%% file: main.tex
\documentclass[10pt,proc]{article} 

\usepackage{booktabs}
\usepackage{graphicx}
\usepackage{xcolor}

\usepackage{multirow}
\usepackage[affil-it]{authblk}
\usepackage{amsmath}
\usepackage{url}
\usepackage{fullpage}

\usepackage{etoolbox}
\patchcmd{\paragraph}{\@parfont}{\bfseries}{}{}
\patchcmd{\paragraph}{\parindent}{0pt}{}{}

\begin{document}

\title{Comparing Video Based Shoulder Surfing with Live Simulation\thanks{This article appears in the proceedings of the 2018 Annual Computer Security Applications Conference.}}

\author{
Adam J. Aviv${}^\bullet$\thanks{Corresponding Author} , Flynn Wolf ${}^\diamond$, and Ravi Kuber ${}^\diamond$\\
\small
${}^\bullet$ United States Naval Academy\\
${}^\diamond$ University of Maryland,Baltimore County\\
{\tt aviv@unsa.edu} {\tt \{flynn.wolf,rkuber\}@umbc.edu}
}

%
%




\maketitle

\input{abstract}

\input{tables} 

\input{intro}

\input{related}

\input{methods}

\input{limits}

\input{results}

\input{posthoc}

\input{takeawaysandimplications}

\input{conclusion}

\section*{Acknowledgments}
This work was supported by the Office of Naval Research.  The authors wish to thank Chukwuemeka KC Marume and John T. Davin for their assistance conducting the study. 

\clearpage
\bibliographystyle{acm}
\bibliography{ref} 

\appendix
\input{appendix_FMW}
\input{pattern_viz}

\end{document}

%% file: abstract.tex
\begin{abstract}
  We analyze the claims that video recreations of shoulder surfing attacks offer
  a suitable alternative and a baseline, as compared to evaluation in a live
  setting.  We recreated a subset of the factors of a prior video-simulation
  experiment conducted by Aviv et al. (ACSAC 2017), and model the same
  scenario using live participants ($n=36$) instead (i.e., the victim and
  attacker were both present).  The live experiment confirmed
  that for Android's graphical patterns video simulation is consistent with the
  live setting for attacker success rates. However, both 4- and 6-digit PINs
  demonstrate statistically significant differences in attacker performance,
  with live attackers performing as much 1.9x better than in the video
  simulation. The security benefits gained from removing feedback lines in
  Android's graphical patterns are also greatly diminished in the live setting,
  particularly under multiple attacker observations, but overall, the data
  suggests that video recreations can provide a suitable baseline measure for
  attacker success rate. However, we caution that researchers should consider
  that these baselines may greatly underestimate the threat of an attacker in
  live settings.
\end{abstract}


%% file: tables.tex
\newcommand{\tableviews}{
\begin{table*}
 \resizebox{\linewidth}{!}{
   \begin{tabular}{c | c | c | c | c | c | c | c }     
         \multicolumn{2}{c|}{}    &    \multicolumn{2}{c|}{\em One-View}  &      \multicolumn{2}{c|}{\em Two-Same}   & \multicolumn{2}{c}{\em Two-Different}  \\
     {\bf Auth.} & {\bf Length}   &    {\bf Live} & {\bf Video}       &      {\bf Live} & {\bf Video}        & {\bf Live} & {\bf Video}           \\
\hline
\multirow{4}{*}{PAT}& \multirow{2}{*}{4-len} & 50/53=94.3\% & 106/111=95.5\% & 15/17=88.2\% & 24/27=88.9\% & 19/20=95.0\% & 62/68=91.2\%\\
\cline{3-8}
 & & \multicolumn{2}{c|}{$\chi^2=0.00,p=1.00,\delta_{95}[-0.10,0.07]$} & \multicolumn{2}{c|}{$\chi^2=0.00,p=1.00,\delta_{95}[-0.21,0.19]$} & \multicolumn{2}{c}{$\chi^2=0.01,p=0.93,\delta_{95}[-0.11,0.19]$}\\
\cline{2-8}
& \multirow{2}{*}{6-len} & 45/55=81.8\% & 74/95=77.9\% & 18/19=94.7\% & 23/24=95.8\% & 14/16=87.5\% & 78/84=92.9\%\\
\cline{3-8}
 & & \multicolumn{2}{c|}{$\chi^2=0.13,p=0.72,\delta_{95}[-0.11,0.19]$} & \multicolumn{2}{c|}{$\chi^2=0.00,p=1.00,\delta_{95}[-0.15,0.13]$} & \multicolumn{2}{c}{$\chi^2=0.05,p=0.82,\delta_{95}[-0.26,0.15]$}\\
\cline{2-8}
\hline
\multirow{4}{*}{NPAT}& \multirow{2}{*}{4-len} & {\bf 48/55=87.3\%} & {\bf 74/104=71.2\%} & 18/19=94.7\% & 17/20=85.0\% & 16/16=100.0\% & 49/61=80.3\%\\
\cline{3-8}
 & & \multicolumn{2}{c|}{$\chi^2=4.37,p=0.04*,\delta_{95}[0.02,0.30]$} & \multicolumn{2}{c|}{$\chi^2=0.22,p=0.64,\delta_{95}[-0.14,0.33]$} & \multicolumn{2}{c}{$\chi^2=2.38,p=0.12,\delta_{95}[0.06,0.34]$}\\
\cline{2-8}
& \multirow{2}{*}{6-len} & 35/53=66.0\% & 58/101=57.4\% & {\bf 17/17=100.0\%} & {\bf 15/24=62.5\%} & 18/20=90.0\% & 56/78=71.8\%\\
\cline{3-8}
 & & \multicolumn{2}{c|}{$\chi^2=0.75,p=0.39,\delta_{95}[-0.09,0.26]$} & \multicolumn{2}{c|}{$\chi^2=6.13,p=0.01*,\delta_{95}[0.13,0.62]$} & \multicolumn{2}{c}{$\chi^2=1.95,p=0.16,\delta_{95}[-0.01,0.38]$}\\
\cline{2-8}
\hline
\multirow{4}{*}{PIN}& \multirow{2}{*}{4-len} & {\bf 89/111=80.2\%} & {\bf 48/94=51.1\%} & {\bf 32/33=97.0\%} & {\bf 12/21=57.1\%} & {\bf 34/36=94.4\%} & {\bf 37/65=56.9\%}\\
\cline{3-8}
 & & \multicolumn{2}{c|}{$\chi^2=18.17,p=0.00*,\delta_{95}[0.16,0.43]$} & \multicolumn{2}{c|}{$\chi^2=10.98,p=0.00*,\delta_{95}[0.14,0.66]$} & \multicolumn{2}{c}{$\chi^2=13.88,p=0.00*,\delta_{95}[0.21,0.54]$}\\
\cline{2-8}
& \multirow{2}{*}{6-len} & {\bf 46/105=43.8\%} & {\bf 17/109=15.6\%} & {\bf 25/39=64.1\%} & {\bf 4/17=23.5\%} & {\bf 31/36=86.1\%} & {\bf 19/68=27.9\%}\\
\cline{3-8}
 & & \multicolumn{2}{c|}{$\chi^2=19.16,p=0.00*,\delta_{95}[0.16,0.41]$} & \multicolumn{2}{c|}{$\chi^2=6.27,p=0.01*,\delta_{95}[0.11,0.70]$} & \multicolumn{2}{c}{$\chi^2=29.62,p=0.00*,\delta_{95}[0.41,0.76]$}\\
\cline{2-8}
\hline
   \end{tabular}}
 \caption{Attacker accuracy results for the live experiment and the video experiment. The view type indicates if the participant provides a single (or one) view or multiple views (two), either from the same angle or different angles of observation. For the video study, only data where screen base resolution $>1800$ with left or right views (no top) was considered. A 2-sample test for equality of proportions with continuity correction was used, and the $\chi^2$ statistic,  $p$-value, and 95\% confidence interval ($ \delta_{95}$) of the difference between the proportions ({\em live} - {\em video}) are reported.}
 \label{tab:views}
\end{table*}%
}

\newcommand{\tableangles}{
\begin{table*}
  \centering
\small
    \begin{tabular}{c | c | c | c | c | c }
      \multicolumn{2}{c|}{}                 & \multicolumn{2}{c|}{\em Left} & \multicolumn{2}{c}{\em Right}  \\
      {\bf Auth.} & {\bf Length}   & {\bf Live} & {\bf Video}  & {\bf Live} & {\bf Video}  \\
      \hline
\multirow{4}{*}{PAT}& \multirow{2}{*}{4-len} & 27/28=96.4\% & 47/49=95.9\% & 23/25=92.0\% & 59/62=95.2\%\\
\cline{3-6}
 & & \multicolumn{2}{c|}{$\chi^2=0.00,p=1.00,\delta_{95}[-0.09,0.10]$} & \multicolumn{2}{c}{$\chi^2=0.00,p=0.95,\delta_{95}[-0.18,0.12]$}\\
\cline{2-6}
& \multirow{2}{*}{6-len} & 22/26=84.6\% & 36/48=75.0\% & 23/29=79.3\% & 38/47=80.9\%\\
\cline{3-6}
 & & \multicolumn{2}{c|}{$\chi^2=0.44,p=0.51,\delta_{95}[-0.12,0.31]$} & \multicolumn{2}{c}{$\chi^2=0.00,p=1.00,\delta_{95}[-0.22,0.19]$}\\
\cline{2-6}
\hline
\multirow{4}{*}{NPAT}& \multirow{2}{*}{4-len} & 23/29=79.3\% & 34/46=73.9\% & {\bf 25/26=96.2\%} & {\bf 40/58=69.0\%}\\
\cline{3-6}
 & & \multicolumn{2}{c|}{$\chi^2=0.07,p=0.80,\delta_{95}[-0.17,0.28]$} & \multicolumn{2}{c}{$\chi^2=6.11,p=0.01*,\delta_{95}[0.10,0.44]$}\\
\cline{2-6}
& \multirow{2}{*}{6-len} & 15/25=60.0\% & 29/52=55.8\% & 20/28=71.4\% & 29/49=59.2\%\\
\cline{3-6}
 & & \multicolumn{2}{c|}{$\chi^2=0.01,p=0.92,\delta_{95}[-0.22,0.31]$} & \multicolumn{2}{c}{$\chi^2=0.69,p=0.41,\delta_{95}[-0.12,0.37]$}\\
\cline{2-6}
\hline
\multirow{4}{*}{PIN}& \multirow{2}{*}{4-len} & {\bf 44/54=81.5\%} & {\bf 22/45=48.9\%} & {\bf 45/57=78.9\%} & {\bf 26/49=53.1\%} \\
\cline{3-6}
 & & \multicolumn{2}{c|}{$\chi^2=10.31,p=0.00*,\delta_{95}[0.13,0.53]$} & \multicolumn{2}{c}{$\chi^2=6.86,p=0.01*,\delta_{95}[0.06,0.45]$}\\
\cline{2-6}
& \multirow{2}{*}{6-len} & {\bf 25/54=46.3\%} & {\bf 6/45=13.3\%} & {\bf 21/51=41.2\%} & {\bf 11/64=17.2\%}\\
\cline{3-6}
 & & \multicolumn{2}{c|}{$\chi^2=10.91,p=0.00*,\delta_{95}[0.14,0.52]$} & \multicolumn{2}{c}{$\chi^2=6.98,p=0.01*,\delta_{95}[0.06,0.42]$}\\
\cline{2-6}
\hline
\end{tabular}

\caption{Effects on angle on attacker accuracy. The angle is either an observation from the left or right with a single view (no repeat viewings). For video-based results, no top views were considered. The prior ``far'' type angles for each side are reduced to simply, left or right, and only data where screen base resolution $>1800$ was considered. A 2-sample test for equality of proportions with continuity correction was used, and the $\chi^2$ statistic,  $p$-value, and 95\% confidence interval ($ \delta_{95}$) of the difference between the proportions ({\em live} - {\em video}) are reported.}
\label{tab:angles}
\end{table*}}

\newcommand{\tablecodes}{
\begin{table}
\centering
\begin{tabular}{c |  r |  c || c | c}
 & {\bf Passcode} & {\bf Live} & {\bf Video} &  \\
  \hline
  \multirow{10}{*}{\rotatebox{90}{PAT}} 
& 0145 & 8/10=80.0\% & 23/23=100.0\% & $\chi^2=2.01,p=0.16,\delta_{95}[-0.52,0.12]$\\
& 1346 & 11/11=100.0\% & 26/28=92.9\% & $\chi^2=0.01,p=0.92,\delta_{95}[-0.09,0.23]$\\
& 3157 & 10/10=100.0\% & 28/29=96.6\% & $\chi^2=0.00,p=1.00,\delta_{95}[-0.07,0.14]$\\
& 4572 & 11/11=100.0\% & 17/19=89.5\% & $\chi^2=0.13,p=0.72,\delta_{95}[-0.10,0.32]$\\
& 6745 & 10/11=90.9\% & 12/12=100.0\% & $\chi^2=0.00,p=0.96,\delta_{95}[-0.35,0.17]$\\
\cline{2-5}
& 014763 & 8/10=80.0\% & 15/19=78.9\% & $\chi^2=0.00,p=1.00,\delta_{95}[-0.31,0.33]$\\
& 136785 & 11/11=100.0\% & 14/17=82.4\% & $\chi^2=0.72,p=0.40,\delta_{95}[-0.08,0.43]$\\
& 642580 & 9/9=100.0\% & 21/22=95.5\% & $\chi^2=0.00,p=1.00,\delta_{95}[-0.09,0.18]$\\
& 743521 & 5/12=41.7\% & 11/21=52.4\% & $\chi^2=0.05,p=0.82,\delta_{95}[-0.52,0.31]$\\
& 841257 & 12/13=92.3\% & 13/16=81.2\% & $\chi^2=0.10,p=0.75,\delta_{95}[-0.20,0.42]$\\
\hline\hline
 
  \multicolumn{5}{c}{} \\ \hline\hline
\multirow{10}{*}{\rotatebox{90}{NPAT}} 
& 0145 & 10/11=90.9\% & 12/16=75.0\% & $\chi^2=0.29,p=0.59,\delta_{95}[-0.19,0.51]$\\
& 1346 & 8/10=80.0\% & 14/18=77.8\% & $\chi^2=0.00,p=1.00,\delta_{95}[-0.31,0.36]$\\
& 3157 & 11/11=100.0\% & 15/21=71.4\% & $\chi^2=2.22,p=0.14,\delta_{95}[0.02,0.55]$\\
& 4572 & 7/10=70.0\% & 12/25=48.0\% & $\chi^2=0.65,p=0.42,\delta_{95}[-0.19,0.63]$\\
& 6745 & 12/13=92.3\% & 21/24=87.5\% & $\chi^2=0.00,p=1.00,\delta_{95}[-0.20,0.29]$\\
\cline{2-5}
& 014763 & 12/14=85.7\% & 15/21=71.4\% & $\chi^2=0.33,p=0.57,\delta_{95}[-0.18,0.47]$\\
& 136785 & 5/10=50.0\% & 7/19=36.8\% & $\chi^2=0.08,p=0.77,\delta_{95}[-0.32,0.59]$\\
& 642580 & 7/9=77.8\% & 16/26=61.5\% & $\chi^2=0.23,p=0.63,\delta_{95}[-0.24,0.57]$\\
& 743521 & 3/9=33.3\% & 12/18=66.7\% & $\chi^2=1.52,p=0.22,\delta_{95}[-0.79,0.13]$\\
& 841257 & 8/11=72.7\% & 8/17=47.1\% & $\chi^2=0.90,p=0.34,\delta_{95}[-0.17,0.69]$\\
\hline\hline

  \multicolumn{5}{c}{} \\ \hline\hline
\multirow{10}{*}{\rotatebox{90}{PIN}} 
& 1328 & 15/21=71.4\% & 16/28=57.1\% & $\chi^2=0.53,p=0.47,\delta_{95}[-0.17,0.45]$\\
& 1955 & 18/21=85.7\% & 11/19=57.9\% & $\chi^2=2.60,p=0.11,\delta_{95}[-0.04,0.60]$\\
& {\bf 5962} & 22/24=91.7\% & 6/18=33.3\% & $\chi^2=13.23,p=0.00*,\delta_{95}[0.29,0.88]$\\
& 6702 & 15/21=71.4\% & 6/17=35.3\% & $\chi^2=3.61,p=0.06,\delta_{95}[0.01,0.71]$\\
& 7272 & 19/24=79.2\% & 9/12=75.0\% & $\chi^2=0.00,p=1.00,\delta_{95}[-0.29,0.38]$\\
& 1328 & 15/21=71.4\% & 16/28=57.1\% & $\chi^2=0.53,p=0.47,\delta_{95}[-0.17,0.45]$\\
\cline{2-5}
& 153525 & 8/24=33.3\% & 4/13=30.8\% & $\chi^2=0.00,p=1.00,\delta_{95}[-0.31,0.37]$\\
& {\bf 159428} & 12/21=57.1\% & 1/23=4.3\% & $\chi^2=12.27,p=0.00*,\delta_{95}[0.25,0.80]$\\
& {\bf 366792} & 8/21=38.1\% & 2/26=7.7\% & $\chi^2=4.72,p=0.03*,\delta_{95}[0.03,0.58]$\\
& 441791 & 10/21=47.6\% & 5/27=18.5\% & $\chi^2=3.40,p=0.07,\delta_{95}[-0.01,0.59]$\\
& 458090 & 8/18=44.4\% & 5/20=25.0\% & $\chi^2=0.84,p=0.36,\delta_{95}[-0.16,0.55]$\\
\hline
\hline  
\end{tabular}
\caption{Comparison of attacker accuracy per-passcode. Only single view conditions were considered in both live and video, and for the video results, only data with resolution $>1800$ was included and the ``top'' angle was excluded. A 2-sample test for equality of proportions with continuity correction was used, and the $\chi^2$ statistic,  $p$-value, and 95\% confidence interval ($ \delta_{95}$) of the difference between the proportions ({\em live} - {\em video}) are reported.}
\label{tab:codes}
\end{table}
}


%% file: intro.tex
\section{Introduction}


Biometric authentication mechanisms offer considerable promise to smartphone
users. However, the protection of unlock authentication still relies on choosing
hard to guess passcodes (e.g., PINs and unlock patterns), while not revealing
those passcodes to untrusted parties.  A common means of attack for gaining access to the
passcode is via {\em shoulder surfing}. In a shoulder surfing attack, an
observer attempts to view a victim in the process of entering his/her passcode
with the intention of recreating that passcode after gaining possession of the
device~\cite{wiedenbeck2006design}.

The area of shoulder surfing has been the subject of a great deal of work~\cite{egelman2014areyouready, forget2010eyegaze, man2003shoulder,
  deluca2014nowyouseeme, deluca2009look, deluca2012touchme, deluca2010colorpin,
  kumar2007reducing,gao2010new,aviv2017shoulder}, for both understanding the
threat and proposing mechanisms to prevent it. Of particular relevance to this
study (termed "current study"), is the work conducted by Aviv et al.~\cite{aviv2017shoulder} (termed: "prior study").  The prior study examined the shoulder surfing susceptibility of three commonly used unlock authentication mechanisms: 4- and 6-digit PINs, 4- and 6-length Android
graphical patterns, and 4- and 6-length Android graphical patterns with the
feedback display turned off (lines rendered by the interface between grid points as they are touched by the user). Due to the difficult nature of evaluating shoulder
surfing attacks in the field, the goal of the prior study was to establish
baselines for shoulder surfing vulnerability in controlled settings that can be
used to compare across authentication types and used as baseline for evaluating
authentication systems that are designed to defend against such attacks.

To control the analysis, the prior study was conducted using a video-based
methodology where the researchers recorded a set of videos with highly
controlled factors and then asked participants to view these videos as a
simulated shoulder surfing scenario.  The data was analyzed to determine
shoulder-surfing susceptibility under each condition. The attack rate (how effectively the participant could recall the passcode entered in the video) was the primary metric.

In this paper, we seek to compare the video-based methodology to a similarly
controlled live setting. In particular, we are interested in assessing the prior
work's following findings relating to the attack success rate.

\begin{itemize}
\item Longer authentication lengths (e.g, 4-digit vs. 6-digit PINs) are less
  vulnerable.
\item PIN authentication is less vulnerable to the attack compared to patterns
  with and without feedback lines.
\item Removing the feedback lines from patterns decreases the vulnerability to
  shoulder surfing.
\item Multiple observations increases vulnerability.
\item Video based evaluation provides a baseline for live, in-person shoulder
  surfing vulnerability.
\end{itemize}


Using the raw results of the prior study, we compare the attacker success rates
of the live setting to a comparable subset of the video study data. Testing for
differences in proportionality, we are {\em unable} to reject the null
hypothesis that the attacker success rate are the same for Android patterns as
well as in many of the settings with patterns without feedback lines. This
suggests that there is consistency between the results of the video and live
simulations. However, the advantage of removing feedback lines previously
observed in video simulation is considerably lessened in the live setting.  For
PINs, we observe significant difference between the video and the live settings,
where live attackers performed up to 1.9x better in some scenarios. Stereo
vision seems to greatly improve the reliability of recalling the more complex
motions of entering a PIN. Despite this discrepancy, the claim of Aviv et al. of
these results forming a baseline is still supported: we never observed a
situation by which the live simulation performed worse than a video study when
significant differences exist.




We conclude that video studies do provide a reasonable approximation for live
simulation of shoulder surfing in settings that involve graphical passwords (but
not PINs), like the Android password pattern, and at least a lower-bound on the
attack success rate for all tested authentication types (including PINs).
However, researchers should consider that this lower-bound may be a significant
underestimation compared to the true attack rate in live simulations. 

\vspace{5 mm}


%% file: related.tex
\section{Related Work}
\paragraph{Mobile authentication and observation attacks} 
Threats such as shoulder-surfing attacks 
have been well documented by researchers \cite{wiese2015pitfalls,aviv2017shoulder}. Studies
have been conducted examining experiences of users who had encountered
observation attacks~\cite{eiband2017understanding} where shoulder surfing was
found to be ``casual'' and ``opportunistic.'' Harbach et
al. \cite{harbach2014hardlock} found that participants only very rarely
reported shoulder surfing (0.3\% of 1134 sampled events) as an immediate high
risk threat when authenticating.

In order to minimize the risk associated with observation attacks, users are
known to modify their own usage behaviors when using a mobile device, hiding the
device from sight and performing mobile interactions in the pocket or bag, or
even shielding the screen \cite{abdolrahmani2016empirical}. Solutions also exist
to obscure screens from third parties \cite{deluca2014nowyouseeme}, to detect
the presence of shoulder surfers in a nearby vicinity \cite{ryu2017} or to
deceive onlookers from data being entered \cite{von2015swipin,
  krombholz2016force}.  Attacks have also been simulated by having observers watch video
footage of victims entering authentication sequences.  Examples include \cite{khan2018CHI} where attacks took place from top and side views. A range of solutions have also been proposed to minimize the likelihood of shoulder-surfing when entering authentication
sequences \cite{ali2016developing}.  However, as highlighted by Wiese and Roth
\cite{wiese2015pitfalls}, it can be difficult to compare the efficacy of these
solutions, as the ways in which these systems are studied varies. Furthermore,
the outcomes can be difficult to compare and interpret.

\paragraph{Evaluating resistance from shoulder surfing}
Many evaluation studies have focused on observing unlock screen interactions
where PINs and patterns are entered
\cite{schaub2012password,aviv2017shoulder,khan2018CHI}. Wiese and Roth
\cite{wiese2015pitfalls} suggest that conducting such studies are challenging
because real-world adversaries are not available for study and must be simulated in one way
or another. In contrast to live studies where participants and
actors/researchers perform tasks together in person, video simulations have been
used to identify susceptibility of on-screen
threats~\cite{sahami2012assessing,ali2016developing}. Video recordings offer
consistency when presented to multiple users \cite{wiese2015pitfalls}, and can
also be accessed independent of location. However, research indicates that that
the success of adversaries is lower when performing video observations compared to live
settings \cite{schaub2013exploring, wiese2015pitfalls}; we make a similar
observation here. Prior research also recommends that shoulder surfing attackers
should be allowed a number of observations~\cite{wiese2015pitfalls} as well as
viewing interactions from a range of
views~\cite{sahami2012assessing,aviv2017shoulder} and different properties of
passcodes~\cite{aviv2017shoulder}. Additionally, the hand position
\cite{schaub2012password} and interaction style when entering data into the
device \cite{aviv2017shoulder} should also be considered. We tested scenarios found to be significant in Aviv et al., following similar procedures.

\paragraph{Overview of Aviv et al.~\cite{aviv2017shoulder}}
Aviv et al. considered the lack of a baseline for comparing common unlock
authentication mechanisms under the threat of shoulder surfing. As a method of
creating such a baseline, the authors used a series of controlled video
simulations of a victim entering unlock authentications using several methods. These methods were PINs and
Android's graphical pattern unlock, with and without feedback lines
present. Additional factors were considered, including the angle of observation,
number of observations, the number of recreation attempts by the observer, the
hand posture of the victim, phone size, and spatial layout of the passcodes.

The methodology of that experiment was multi-factorial. Participants
were selected into one of a number of independent factors (phone type, passcode
choice, authentication type, hand posture) and then a set of randomized
dependent factors (passcodes, observation angles, number of views, and
attempts). For recruitment, the primary results were based off participants
on Amazon Mechanical Turk ($n=1173$) and participants recruited locally
($n=91$), with both groups completing a web survey whereby they viewed videos of
authentication and attempted to recreate the passcodes observed.

Using the results, the authors tested the following hypotheses (the {\bf -p}
indicates a prior work hypothesis):
\begin{itemize}
\item {\bf H1-p}: The type of unlock authentication, PIN pattern with lines,
  patterns without lines, affects the shoulder surfing vulnerability.
\item {\bf H2-p}: Repeated viewing of user input increases the likelihood of a
  shoulder surfing vulnerability.
\item {\bf H3-p}: Multiple attempts to recreate the input affects the likelihood
  of a shoulder surfing vulnerability.
\item {\bf H4-p}: The angle of observations affects shoulder surfing vulnerability.
\item {\bf H5-p}: The properties of the unlock authentication, such as length and
  visual features, affect shoulder surfing vulnerability.
\item {\bf H6-p}: The phone size affects shoulder surfing vulnerability.
\item {\bf H7-p}: The hand position used to hold and interact with a device
  affects shoulder surfing vulnerability.
\end{itemize}
Of those hypotheses, {\bf H1-p}, {\bf H2-p}, {\bf H3-p}, {\bf H4-p}, and {\bf H6-p} were accepted, while {\bf H5-p} was partially accepted, and {\bf H7-p} was
rejected. The authors claim that the video studies, generally, can form a
reasonable replacement for live simulation, and that at the very least a video
study could provide a baseline for shoulder-surfing vulnerability.


%% file: methods.tex
\section{Methodology}
\label{sec:methods}

To investigate the efficacy of video-based recreations for evaluating
observation attacks, we recreated the study conducted by Aviv et
al.~\cite{aviv2017shoulder} with live participants in a controlled lab
environment.  We asked participants to position themselves in similar locations
to where the cameras were positioned in the prior study.  They then attempted to
shoulder-surf a victim (played by a proctor). We varied the type and length of
authentication sequences, observation angle, and number of repeated viewing
attempts, to determine if these factors impact the success of the attacker. The
results were then compared with Aviv et al.'s findings using a comparable subset
of the prior data.  For simplicity of discussion, we refer to the prior work of
Aviv et al. as the {\em video study} and the results here as the {\em live
  study}.

\paragraph{Hypotheses} In particular, we are interested in testing the following
hypothesis related to the efficacy of video based shoulder surfing experiments
as compared to live settings.

\begin{itemize}
\item {\bf H1-r}: Live shoulder surfing confirms accepting prior hypotheses:
  \begin{itemize}
    \item {\bf H1-p}: The authentication type affects shoulder surfing vulnerability
    \item {\bf H2-p}: Repeated viewing affects shoulder surfing vulnerability
    \item {\bf H4-p}: The angle of observation affects shoulder surfing vulnerability
    \item {\bf H5-p}: The properties of the passcodes affects should surfing vulnerability
    \end{itemize}
  \item {\bf H2-r}: Video simulation forms a baseline of performance compared to
    live settings.
\end{itemize}

\subsection{Study Design and Materials}

\begin{figure}[t]
  \centering
  \includegraphics[width=0.45\linewidth]{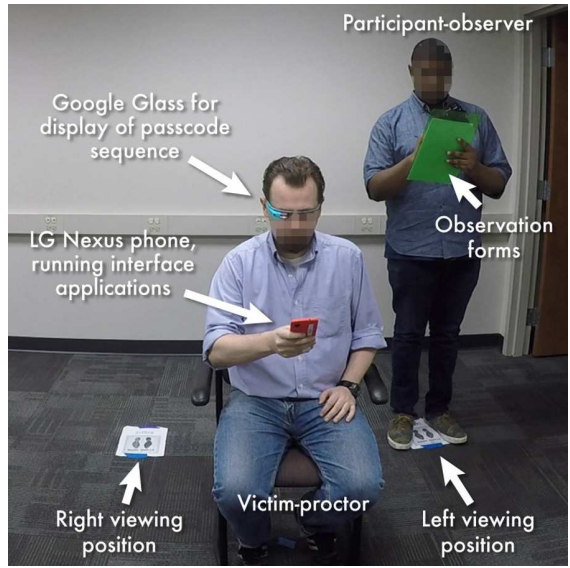}
  \caption{Experimental setup with an observer attacking a victim, a member of a research team. Note the Google Glass displaying the passcode to enter.}
  \label{fig:setup}
\end{figure}

\paragraph{Treatments} 
The study followed a mixed factorial design, similar to the video study.
Independent variables included authentication type (PIN vs pattern) on the Nexus
5 device using the same hand posture/interaction style (one-handed, right thumb
input).  For dependent variables, we reduced the observation angle to two (left
or right) as opposed to the five angles used in prior work. The video study used
the variety of angles to simulate different heights, but height variation is
naturally present in a live study. We kept the same variables for observations
(single observation from one angle, two observations from the same angle, or two
observations from different angles), and we used a lab environment for our live study very similar to the set-up to capture videos for the video study (Aviv et al.) (see
Figure~\ref{fig:setup}).

There were two notable differences between factors in the video study and the
live study. First, we only allowed each participant a single attempt at
recreating the passcode. This choice was motivated by results of the video study
whereby participants, knowing they would have multiple attempts in advance,
actually did worse at the tasks than those that knowingly had one attempt. It
was conjectured that participants attempted to ``game'' the task knowing that
they would have multiple attempts at recreating the passcode. As such, we only
allowed participants to make one recreation attempt, and this fact was
communicated during training.

Another difference in the live study was that passcode recreation occurred
using pen-and-paper, as opposed to a simulation of the device used in the video study. This choice was made to simplify the data collection
procedures for both proctors and participants. 

Finally, as we only tested a subset of the treatments of the prior video study,
we only performed our analytic comparisons on a relevant subset of the video
study data. In particular, we removed data that included a top angle and reduced
the two side angles into a single {\em left} or {\em right}
setting. Additionally, as the video cannot control for monitor display size,
which was a large factor in the prior results, we only used the most ideal
viewing conditions, where the reported y-axis pixels were greater than 1800. We
believe this restriction provided the {\em most fair} comparisons possible given
the potential uncontrolled factors. We discuss limitations and realism further
in Section~\ref{sec:limits}.

\paragraph{Authentication types}
We analyzed three authentication types with two different length settings, as
used in the video study.  These included:
\begin{itemize}
\item {\bf PIN}: 4- or 6-length PINs consisting of a set of numbers.
\item {\bf PAT}: Android unlock patterns consisting of 4 or 6 contact points {\em with} the feedback lines present.
\item {\bf NPAT}: Android unlock patterns consisting of 4 or 6 contact points
  {\em without} the feedback lines present.
\end{itemize}
While the PIN interaction display is as one expects, the presence or absence of grid pattern feedback lines is less well known. When a pattern is entered with feedback lines (PAT), the display will show connecting lines on the screen between grid points touched by the user while entering their passcode shape. Alternatively, the connecting lines are not rendered on screen during passcode entry in the without feedback lines (NPAT) pattern display, although the user must still contact the appropriate points in the correct order. As identified by Aviv et al.~\cite{aviv2017shoulder} and von
Zezschwitz et al.~\cite{vzw2015easy}, the absence of feedback lines can make it more
difficult for an observer to recreate the patterns. As part of {\bf H1-r}, we
will make a similar evaluation.
\begin{table}[t]
\centering
\small
\begin{tabular}{ c | l | l c}
{\bf Auth. id} & {\bf Patterns} & {\bf PINs} \\
\cline{1-3}
0 & 0145   & 1328 & \multirow{10}{*}{\includegraphics[width=0.2\linewidth]{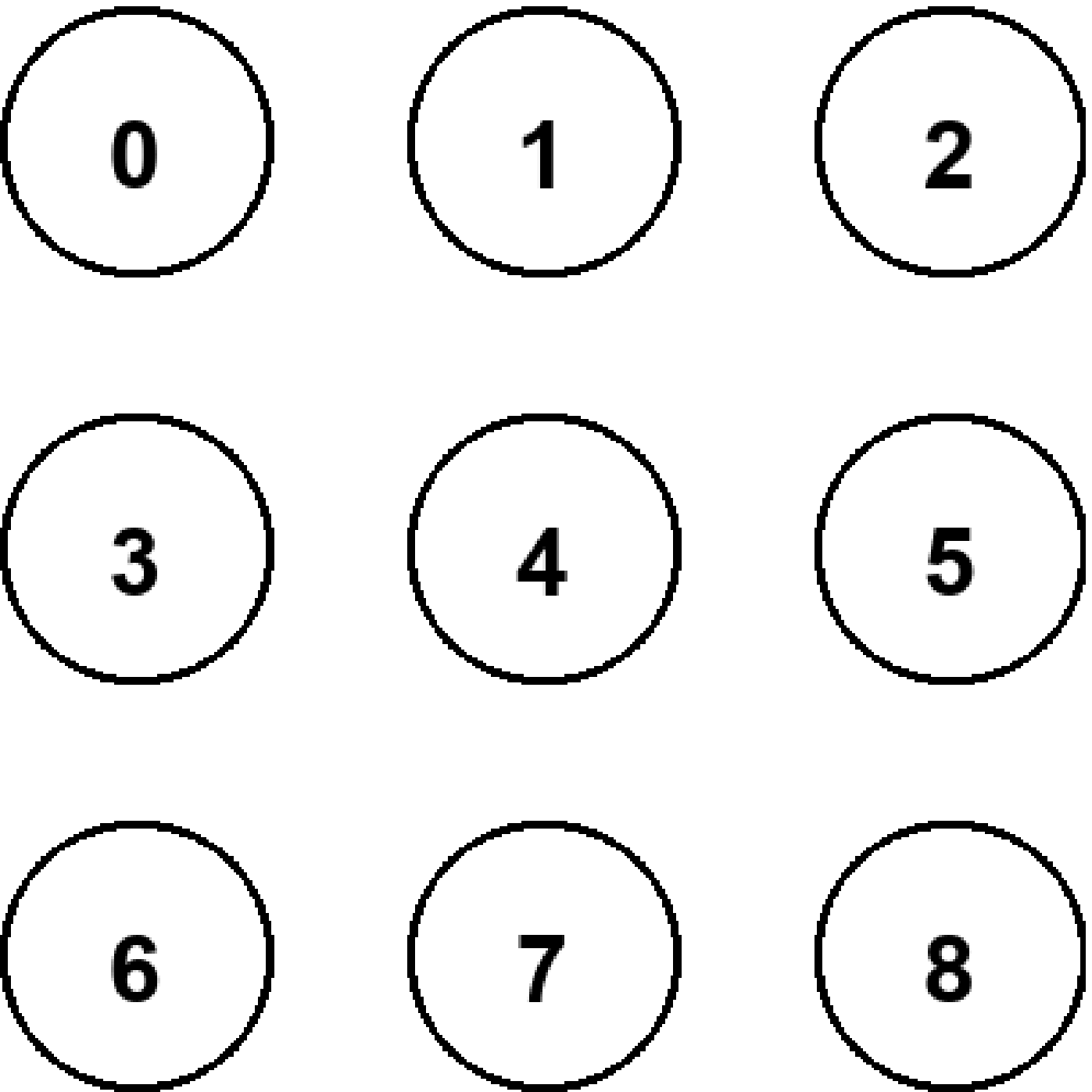}}\\
1 & 014763 & 153525 &\\
2 & 1346   & 159428 &\\
3 & 136785 & 1955 &\\
4 & 3157   & 366792 &\\
5 & 4572   & 441791 &\\
6 & 642580 & 458090 &\\
7 & 6745   & 5962 &\\
8 & 743521 & 6702 &\\
9 & 841257 & 7272 &\\
\end{tabular}
\hfill
\caption{Authentication identifiers for patterns and PINs. To the right, the
  numeric labeling for patterns to contact points.}
\label{tab:passcodes}
\end{table}

To maintain consistency, we used the same set of patterns and PINs as in prior
work (Table~\ref{tab:passcodes} and Appendix \ref{fig:patterns}). The patterns
were selected from an online study of self-reported
patterns~\cite{aviv2015isbigger}, and the PINs were obtained from sequences of
digits in leaked password sets, similar to the analysis by Bonneau et
al.~\cite{bonneau2012birthday}. Further, the set of passcodes were selected for
physical properties, as the layout and sequence of gestures in entry may affect
shoulder surfing attack rate.  The patterns' spatial properties might affect surfing attacks because an attacker's view from some viewing angles might be obscured for some parts of the touchscreen.

\paragraph{Randomization and counterbalancing}
One of the restrictions for performing the study using live participants as
compared to video recreation is that the same level of randomization is nearly
impractical for the target recruitment size and the set of factors being
considered.  As such, we designed a two stage randomization procedure, one for
ordering the passcodes and one for ordering the observation angles.

In particular, Table~\ref{tab:orders} contains three different randomized
orders across the passcode.  These are labeled Order a, b, and c. Note that the
authentication identifiers refer to Table~\ref{tab:passcodes}. In
Table~\ref{tab:angleorder} are four randomized orders for observation angles (i,
ii, iii, and iv). For each participant, we randomly assigned them a 
passcode order and an observation angle, producing 12 different
randomizations. 

At this point, it is important to consider counterbalancing. Selecting
randomized orders for passcodes or observations can weight the data
improperly. This leads to an optimization problem, and we used a utility
function to find a set of randomized orders that would provide (1) sufficient
data in each factor for us to perform statistical tests, (2) a roughly equal ratio of data within each factor being compared (4- vs 6-length,
auth-type, angle), (3) that each passcode only appears once per viewing, and (4) that within
each viewing sequence, per participant, there are roughly an equal number of
single and multiple observations. We found a case that nearly met these
criteria, as displayed in Table~\ref{tab:orders} and~\ref{tab:angleorder}. The
weighting is then displayed based on 12 participants in
Table~\ref{tab:counterbalance}, leaving us with 72 single-view observations and
48 multi-view observations, 24 from the same angle twice and 24 from two different
angles. Additionally, there is equal weighting across angles and viewing
(Table~\ref{tab:angleorder}), and nearly equal weighting across passcodes.

We acknowledge that this counterbalancing is not a perfect weighting, and
solving this particular optimization problem is challenging and may
not have a solution. However, the resulting counterbalancing compares
favorably to the subset of relevant video study data. For PINs, there is nearly an
equal number of observations in the one-view and two-view conditions. For
PAT/NPAT, there is 50\% less observations in one-view condition with a
significant proportion necessary for statistical testing, and the two-view
conditions for PAT/NPAT are of the same magnitude as the video study (see
Table~\ref{tab:views}).


In total, we were able to run complete trials for 18 participants each for PAT and
NPAT, and all of those 36 participants also completed a PIN viewing. The order
between PIN and PAT/NPAT for participants was randomized, so that half of the
participants completed a PIN trial before doing a PAT/NPAT trial, and the other
half completed the protocol in the reverse order, PAT/NPAT then PIN.

\begin{table}[t]
\centering
\small
\begin{tabular}{ c | c c c c c c c c c c}
{\bf Order } & \multicolumn{10}{c}{{\bf Auth. id}}\\
\hline
a & 8& 1& 0& 7& 9& 2& 6& 5& 4& 3\\
b & 0& 6& 3& 8& 2& 4& 9& 7& 1& 5\\
c & 6& 0& 9& 4& 8& 3& 5& 1& 7& 2\\
\end{tabular}
\caption{Orderings of the patterns and PINs in the experiments.}
\label{tab:orders}
\end{table}

\begin{table}[t]
\centering
\footnotesize
\begin{tabular}{ c | c c c c c c c c c c}
{\bf Exp. } & \multicolumn{10}{c}{{\bf Angle(s)}}\\
\hline
i & L & R &  R & RR & L& LR& RL& R& L& LL \\
ii & RL& L& LR& R& LL& R& R& L& L& RR \\
iii &LL& RR& L& R& R& L& R& LR& RL& L \\
iv & LR& R& R& L& L& RL& RR& L& LL& R \\
\end{tabular}%
\caption{Angles used within each experiment, including multiple views with two angles indicated. L=view from left side, R=view from right side.}
\label{tab:angleorder}
\end{table}

\begin{table}[t]
\centering
\small
\begin{tabular}{ c | c c c c c c | c c c}
{\bf Auth. id} & {\bf L} & {\bf R} & {\bf LL} & {\bf RR} & {\bf LR} & {\bf RL} & {\bf one} & {\bf two-same} & {\bf two-different}\\
\hline
  0 & 3 & 4 & 1 & 1 & 2 & 1 & 7 & 2 & 3\\
  1 & 5 & 3 & 1 & 1 & 1 & 1 & 8 & 2 & 2\\
  2 & 4 & 3 & 2 & 1 & 1 & 1 & 7 & 3 & 2\\
  3 & 3 & 4 & 1 & 1 & 2 & 1 & 7 & 2 & 3 \\
  4 & 4 & 3 & 1 & 1 & 1 & 2 & 7 & 2 & 3\\
  5 & 3 & 4 & 1 & 2 & 1 & 1 & 7 & 3 & 2\\
  6 & 2 & 4 & 1 & 2 & 1 & 2 & 6 & 3 & 3\\
  7 & 5 & 3 & 1 & 1 & 1 & 1 & 8 & 2 & 2\\
  8 & 4 & 3 & 2 & 1 & 1 & 1 & 7 & 3 & 2\\
  9 & 3 & 5 & 1 & 1 & 1 & 1 & 8 & 2 & 2\\
  \hline
total   &36&36 & 12 & 12 & 12 & 12 & 72 & 24 & 24
\end{tabular}%
\caption{Balancing of observation angles, number of views, for each authentication after 12 participants, $\text{Order}\times\text{Exp}$.}
\label{tab:counterbalance}
\end{table}


\begin{figure}[t]
  \centering
  \includegraphics[width=0.2\linewidth]{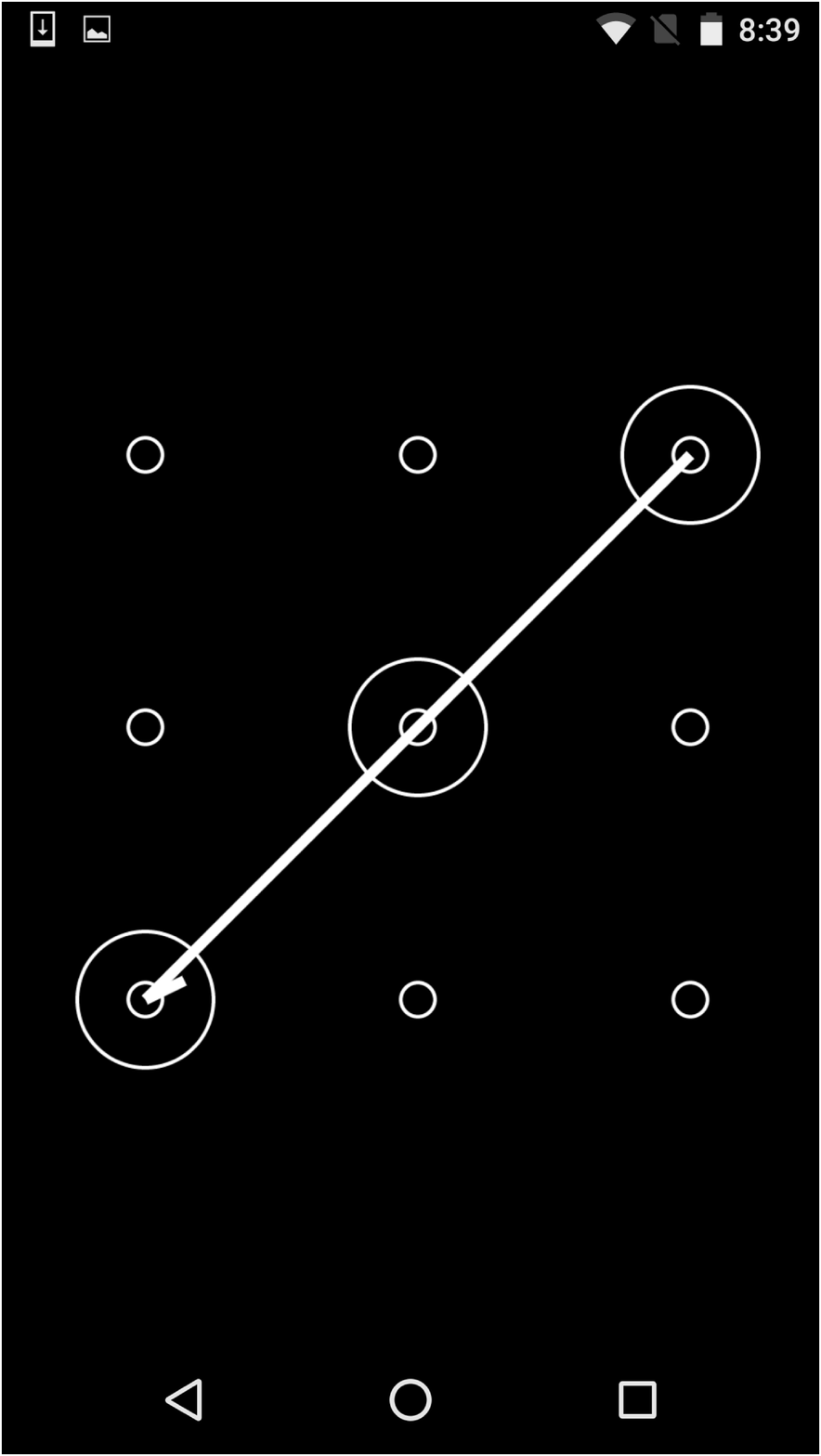}
  \hspace{.2in}
  \includegraphics[width=0.2\linewidth]{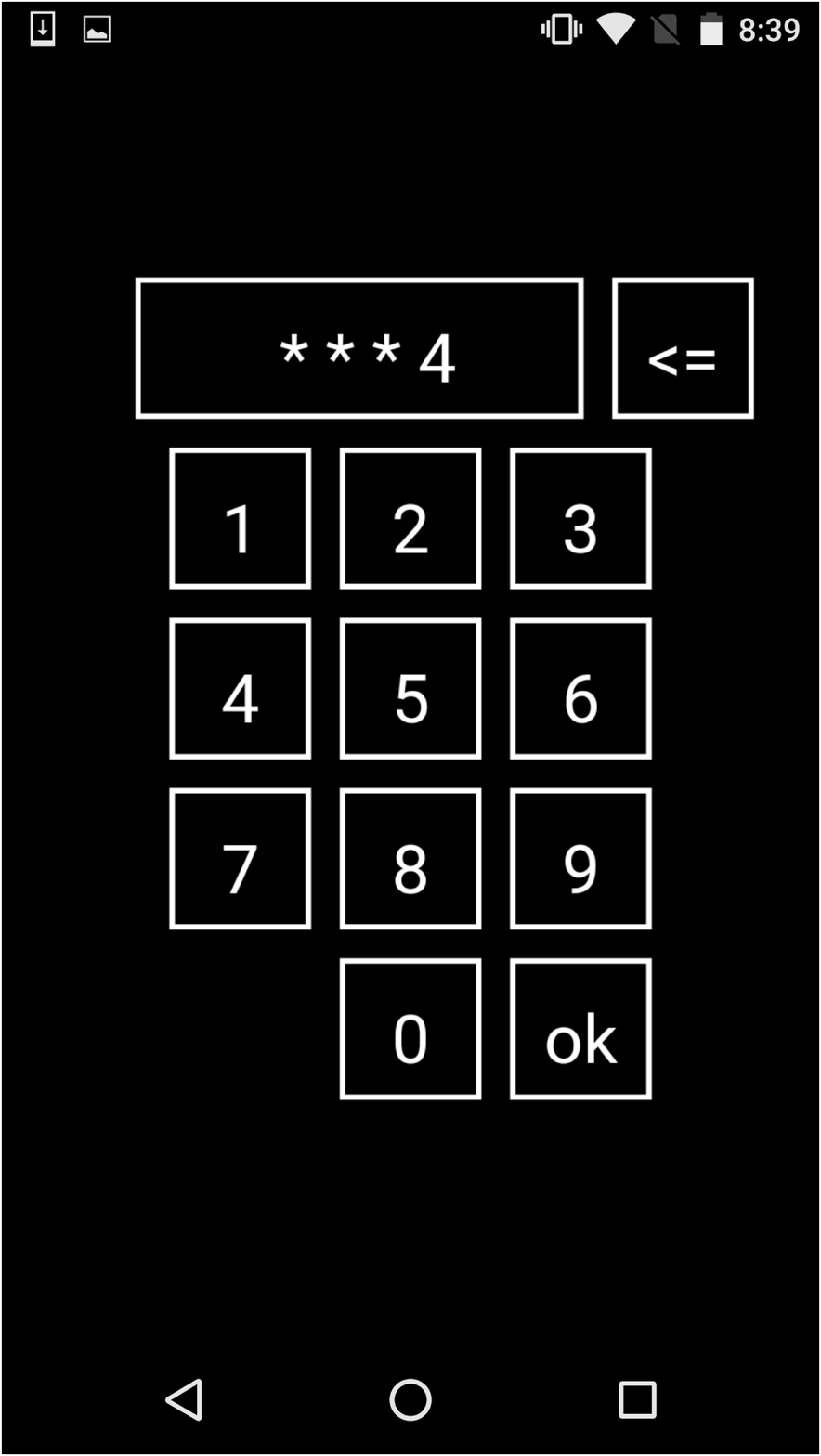}
  \caption{Screenshots of the web-based applications used by the victim entering the passcode. Note, that for the pattern without feedback lines setting, the white trace lines would {\em not} appear.}
  \label{fig:apps}
\end{figure}

\begin{figure}[t]
  \centering
  
  \includegraphics[width=0.45\linewidth]{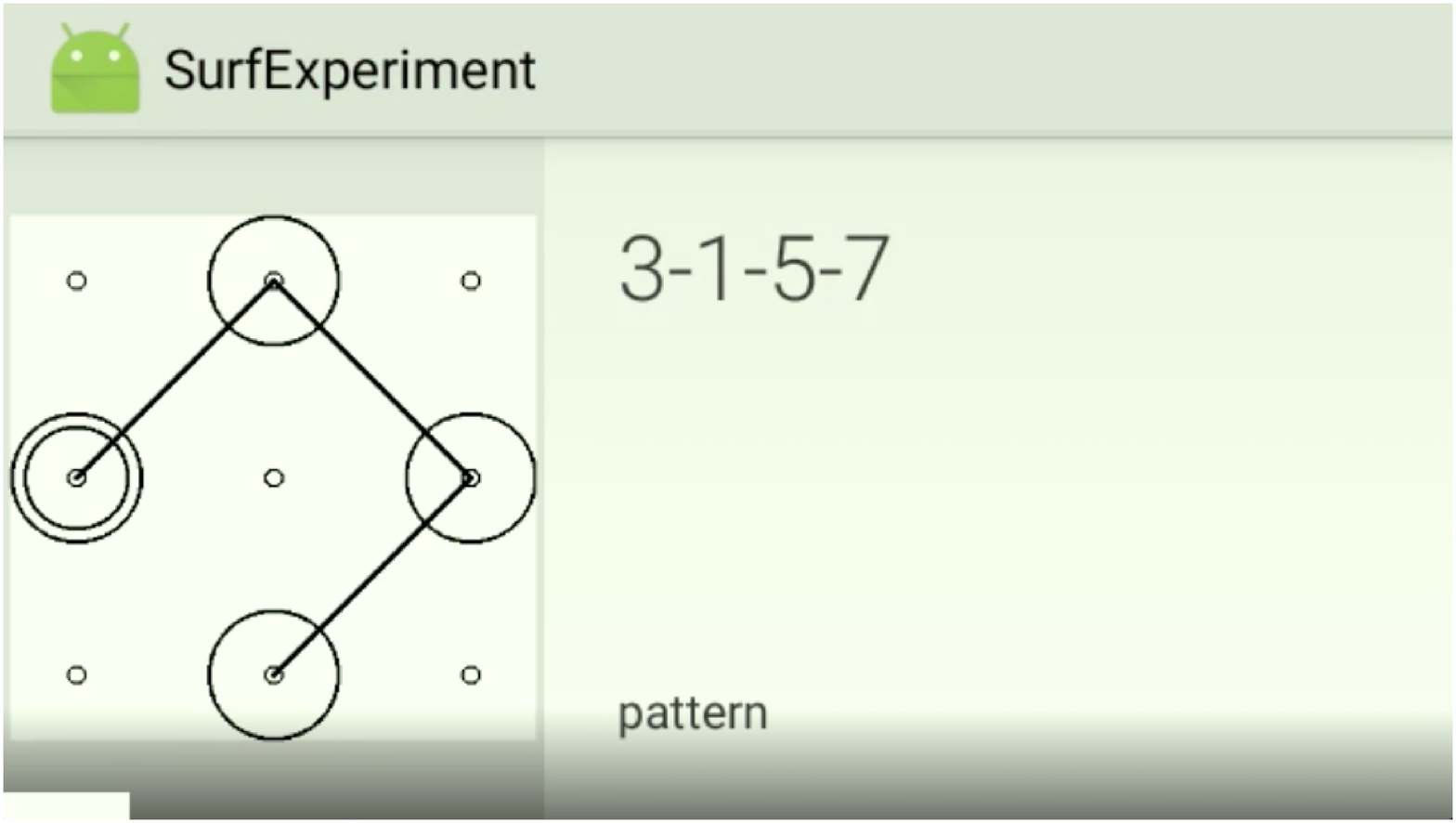}
  \hfill
  \includegraphics[width=0.45\linewidth]{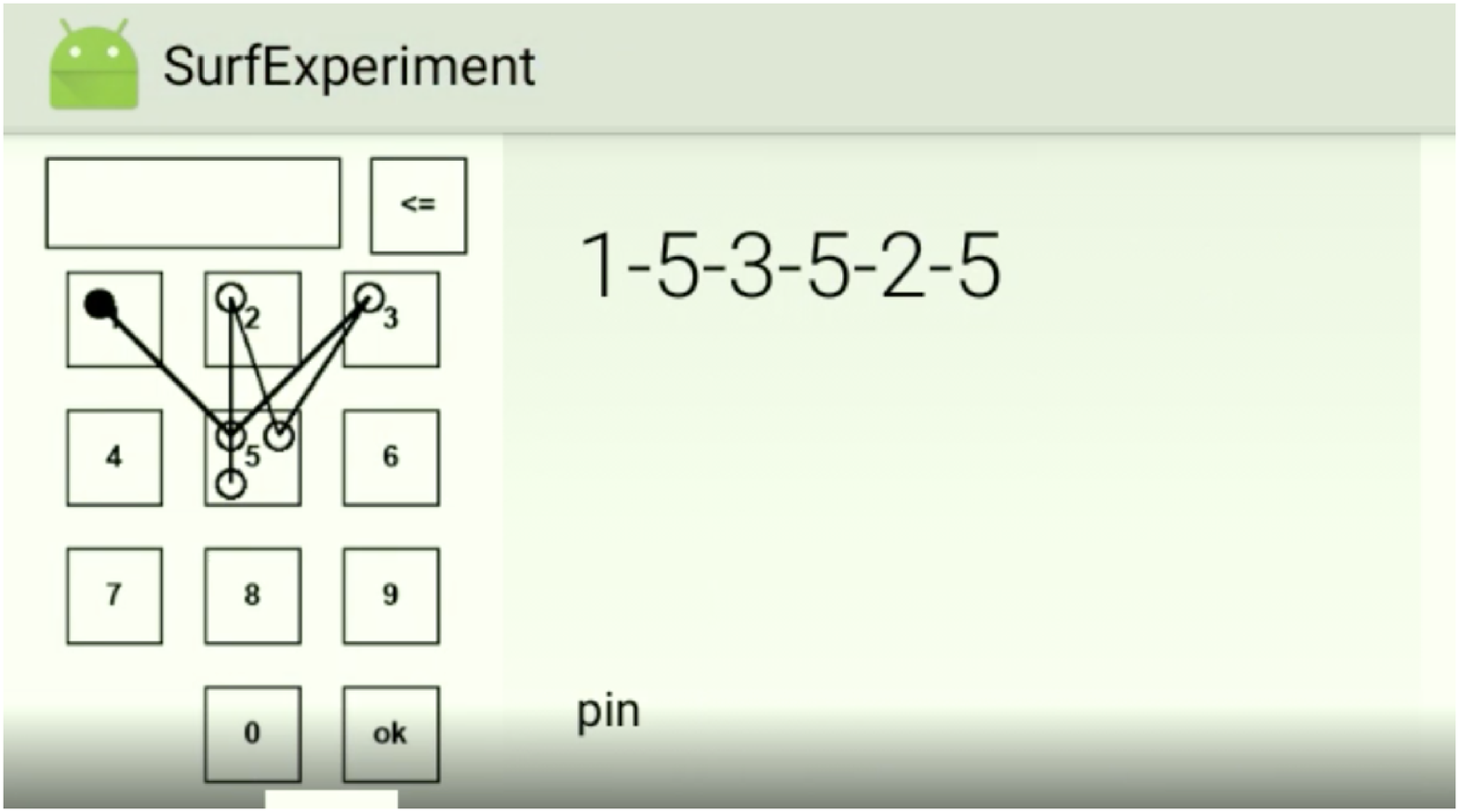}
  \caption{Screenshots of the Google Glass application as viewed by the victim and enter the correct PIN or pattern.}
  \label{fig:glass}
\end{figure}

\subsection{Live Simulation Setup and Coordination}
We sought to recreate nearly the same scenario for shoulder surfing as the video study. Namely, we had our victim placed in a sitting position with the participant observer behind the victim, either standing to the right or the
left, directed by one of two proctors. These were the same positions where the cameras were located (near left and near right views) in the study by Aviv et al.~\cite{aviv2017shoulder}. See Figure~\ref{fig:setup} for a visual of this
arrangement for the live study.  

Additionally, for the phone application used to enter the passcodes, we used the
same mobile applications as in the prior study, which includes a web-based
platform for entering PINs and patterns. Screenshots of those applications are
provided in Figure~\ref{fig:apps}.

For patterns with feedback lines, the white tracing lines would follow the user
gesture, and once the pattern was entered, it would remain visible on-screen for
a half a second before disappearing. The same would be true for the patterns
without feedback lines, however, neither the tracing lines nor the
contact points of the grid would be rendered on the screen. For PINs, the layout
allowed for numeric entry as expected. Once digit keys were selected, the
corresponding digits were presented on the interface.  These would then fade to
a * after a half a second, similar to most mobile PIN entry interfaces.

For the participant observer to record their pattern entry, we used pen and
paper. Examples of the observer forms are provided in the
Appendix (\ref{fig:obvforms}). The forms had text boxes and mini-diagrams of the application interfaces, so
the participants could easily record the observed entry. Participants were asked not to write down the passcodes observed until directed following all observations, which was important for the multiple viewing scenario.

As shown in
Figure~\ref{fig:setup}, two pre-marked spots were placed on the floor to direct
participants where to stand on the left or right side. The second proctor,
following the randomized treatment order, would call out directions to the
participant; for example, ``one view, from the left'' or ``two views, first from
left and then right.'' Once the participant was in place for each view, the
second proctor would cue the first proctor (playing the victim) to enter a passcode.

At this point, a significant challenge we had to overcome was how to prompt the
victim-proctor with the correct passcode to enter without tipping off the
participant-observer. Due to the randomization procedures, requiring the
victim-proctor to memorize the numerous orderings was not realistic. As a
solution, we developed a Google Glass application to guide the victim-proctor
through the various passcode orders. Google Glass is a wearable eyeglass display unit that runs on a modified Android OS. It enables one to scroll interactively
through images projected onto a viewing screen built into the right
eyepiece. Moreover, the small display screen on the Google Glass was not visible to the
participant. A screenshot of the Google Glass application is provided in
Figure~\ref{fig:glass}.  \vspace{5 mm}

\subsection{Procedure}
The replication experiment proceeded in four stages:
\begin{enumerate}
\item Informed Consent and Ante Hoc Questionnaire: All participants were
  properly informed and consented, as we conducted an IRB approved
  experiment. Following consent, we asked participants to complete
  an ante hoc questionnaire that covered basic demographic questions, such as
  age and gender, as well as questions regarding the participants experience
  with smartphones, mobile authentication, and sense of risk from shoulder
  surfing. 
  The subjective response questions were largely intended to orient participants to physical security issues related to the study. The ante hoc questions are found in the Appendix (\ref{app:antehoc}).
\item Training: Depending on the set of authentications being observed in the
  trial run, a training session would include two basic passcodes, the L shape
  for patterns and the 1234 PIN, to help familiarize the participant with the
  procedures, how to record on the observation sheets, and where to stand
  for the trials (similar training was performed in the video study). Additional
  training on how to fill out the observation form was also provided, which is
  included in the Appendix (\ref{fig:obvforms}).
\item Trial: Under the direction of the proctor, the participant conducted 10 observations of either the PAT or NPAT pattern entry, and 10 observations of PIN entry.
\item Post Hoc Questionnaire: Following the trials, the participant answered a
  series of post hoc questions related to the challenge of the task and his/her
  perceived performance thereon. See Appendix~\ref{app:posthoc} for the set of
  post hoc questions.
\end{enumerate}
As each participant completed two trials, one for either PAT or
NPAT and another for PIN, once the trial stage was over for the first
authentication we would return to training for the second authentication.  As a way to control for training effects, whereby observing PINs first could increase
or decrease performance on observing PAT/NPAT, we ensured that there was an even
ratio between the order of the trials. A guide was also followed to ensure that
the researchers followed the same steps in the protocol (see
Appendix~\ref{app:studyguide}).

\begin{table*}[t]
\centering
\small
  \begin{tabular}{ c |r | c | c | c || c | c | c | c}
      & & \multicolumn{3}{c||}{\em Live} & \multicolumn{4}{c}{\em Video} \\
      & & {\bf Male} & {\bf Female} & {\bf Total} & {\bf Male} & {\bf Female} & {\bf Neither} &{\bf Total}\\
\hline
\multirow{6}{*}{\rotatebox{90}{\em Age}} 
&18-24 & 16 & 4 & 10 (27.8\%)  & 30 & 9  & 0 & 39 (39.4\%)\\
&25-34 & 8 & 5 & 13 (36.1\%)   & 24 & 10  & 1 & 34 (34.3\%)\\
&35-44 & 1 & 1 & 2 (5.6\%)     & 10 & 4   & 0 & 14 (14.1\%)\\
&45-54 & 0 & 0 & 0 (0.0\%)     & 4 & 3   & 0 & 7 (7.1\%)\\
&55-64 & 1 & 0 & 1 (2.8\%)     & 1 & 0    & 0 & 1 (1.0\%)\\
& 65+  & 0 & 0 & 0 (0.0\%)     & 1 & 2    & 0 & 3 (3.0\%)\\
\cline{2-9}
&{\bf total} & 26 (72.2\%) & 10 (27.8\%) & {\bf 36} & 70 (70.7\%) & 28 (28.2\%) & 1 (1.0\%) & {\bf 99}\\
\hline
\multirow{3}{*}{\rotatebox{90}{\em Phone}}& iOS & 15 & 7 & 22\\
&Android & 9 & 2 & 11\\
&Windows & 2 & 0 & 2\\
\hline
\multirow{5}{*}{\rotatebox{90}{\em Unlock}}& Fingerprint & 15 & 6 & 21\\
&PIN-6 & 8 & 5 & 13\\
&PIN-4 & 10 & 2 & 12\\
&Pattern & 5 & 1 & 6\\
& None & 3 & 1 & 4
\end{tabular}
\caption{Demographic, phone usage, and unlock authentication types of participants. For the video study, the subset of comparable data that includes participants in both the ``in-person'' and ``online'' settings that had screen resolution greater than 1800px and observed patterns on the Nexus 5 phone.}
\label{tab:dem}
\end{table*}

\subsection{Recruitment}
Participants were recruited from university student mailing lists, and paid \$5
(USD). In total, we recruited 36 participants, including 10 females.  The cohort was predominately aged between 18 to 24 years old. Almost two-thirds of participants used iOS mobile devices. 21 used a fingerprint reader to unlock their phones, and 6 used patterns (we did not ask if feedback lines were turned off).
The demographic breakdown, as well as their choice in mobile device and
authentication are presented in Table~\ref{tab:dem}. 

Additionally presented in Table~\ref{tab:dem} are the demographics of a
comparable set of participants from the prior video study; these participants
observed authentication on the Nexus 5 phone in the ``in-person'' lab setup or
the on-line MTurk setup with a screen resolution of at least 1800px in the
y-axis, the most realistic setting of the prior work. The breakdown of these two
groups are similar, slightly younger overall with about 70/30 gender breakdown.









%% file: limits.tex
\section{Realism and Limitations}
\label{sec:limits}
As described in the previous section, we attempted, as best as possible, to
recreate the settings of the prior video study in live simulation. Due to the complexities of performing such a process, the study described in this paper had its own set of limitations. 


\paragraph{Viewing angles} While we use a similar lab environment for the live
simulation to that used in the video study, the participants could not stand in
exactly the same position as the cameras due to height differences and the
relatively close proximity of the {\em near} and {\em far} angles from a given
side. We thus reduced the observations to simply {\em left} and {\em right} and
relied on the fact that our participants naturally vary in height to compensate
for the {\em near} and {\em far} setting of camera height placement in the prior
study.

\paragraph{Victim entry speed}
Another recreation challenge is that our victim (a proctor)
must enter the authentication sequence many times over at a consistent speed. Clearly, a
video ensures consistency here, and so we trained the victim-proctor on the original
videos to maintain consistent timings of authentication entry. While there is no guarantee that every
participant viewed the authentication at the same rate, we believe this
training, and the total number of entries performed by the victim, ensures
consistency. Further, the same victim-proctor was used in all data collection. 



\paragraph{Subset of conditions} As summarized in Section~\ref{sec:methods}, a
subset of the original conditions were used in the live simulation. We kept
factors that were shown to be significant in the video study, but also had to
remove some that posed usability challenges for the proctor acting as the
victim. While the selection process was done carefully to address conditions
likely to be important, it was also done for a practical nature of conducting a
study with live participants as compared to online. To ensure that we made a
fair comparison, we selected a similar subset of the data from the prior study.
In particular, we used results from the previous study from participants who
had viewing screens of at least 1800px across, who viewed authentication attempts via the Nexus 5 phone with
thumb input from the left or right side.



\paragraph{Pen-and-paper attacker recordings}
As participants were using pen-and-paper to record their observations
during the shoulder surfing attack, some participants were able to use this as
an added aid to support recall of the passcodes. For example, some participants were
viewed by the proctor mimicking the movements made by the victim-proctor between multiple-view conditions
prior to writing down their final observation. While we directed participants to
{\em not} do this during training, it was difficult to stop due to the nature of
the task. In the video study, participants were also directed not to use
additional aids, such as writing down observations while observing the passcodes,
and were required to attest to this.  However, it is possible that the attestations
were not fully truthful, nor could the researchers verify this as the study was
conducted online. As such, as neither study could fully control for this we
believe that this provides for a fair comparison.
\paragraph{Ecological validity}
Low levels of ecological validity are known to be commonplace among lab-based studies for mobile interactions \cite{kjeldskov2014worth}. Although the method and setting selected for our study cannot approximate the conditions by which shoulder surfing may take place in-the-wild, we designed the study to provide a sense of realism even in a lab-based environment (e.g. victim in seated position similar to attacks taking place while seated on public transport, while seated in a classroom, etc.).  However, due to time constraints, conditions such as providing multiple attempts to observe and/or recreate entry, could not be examined. Further study would be needed to widen the range of factors examined, and to identify the applicability of these findings to other types of tasks (e.g. authenticating while ambulatory) or other types of settings (e.g. field-based).  



%% file: results.tex
\tableviews{}

\section{Results}
As the live simulation used a subset of the variables in the video study (see
prior section), we in turn performed comparisons on an appropriate subset of
the video study data. We used video data that met the following criteria:
one-handed/thumb-input on the Nexus 5 (red) phone, viewed from the left or right
angle, and a single recreation attempt. Additionally, we only included video
data that was collected with a screen resolution $>1800$ pixels, which was
identified as the most ideal viewing condition in the prior study
\cite{aviv2017shoulder}. With these reductions, we compared 720 shoulder surfing
attempts for the live simulation to a comparable 1,171 attempts in the video
study.

\subsection{Comparing Attack Rates Across Video and Live Studies}

\paragraph{Statistical Procedures}
As the results of the experiments for both the live and video study are
proportional, either the participant succeeded in recreating the passcode or did
not, we compare the results using a {\em proportionality test} for equality of
proportions, which follows a $\chi^2$ distribution. That is, we compare the attacker success rate for the video study to that of the live study
using the same conditions, reporting the $\chi^2$ statistic, the two-tailed $p$
value, and the 95\% confidence interval ($\delta_{95}$) for the difference
between proportions.

In the cases where $p\le0.05$, we can conclude that the live study was {\em not}
well modeled by the video study because the proportions of attacker success are
significantly different. Similarly if $p>0.05$ we cannot reject the null
hypothesis that the two proportions are the same and thus must conclude that the
proportions are more likely measuring the same effect. The confidence interval
reports the most likely range of difference between the attacker success rate
for the video and live results, but is only relevant when a significant
difference is found.

When comparing data across factors with greater than two conditions, we used a
$\chi^2$ test for goodness of fit to determine significant differences in attack
success rates. Post-hoc analysis is conducted using pairwise comparisons with a
Bonferonni correction.

Across tests, while the data is overlapping for some of the factors being
examined, we do not normalize/correct $p$ values as we are not attempting to
control for type-1 errors across {\em all} tests. Instead, we are performing
exploratory analysis and interested in determining if significant differences
may exist and from where they may arise. In post-hoc analysis, as described
above, we do correct $p$ values as appropriate as this occurs within a single
test with directly overlapping hypothesis.

\paragraph{Authentication Types (H1-r/H1-p)}
In the prior study, a key finding was that a statistical difference was
identified in attacker performance across authentication type. We can perform
the same tests by comparing vulnerability to shoulder surfing for the single
view conditions; see the first column of Table~\ref{tab:views}.

We first compare each of the authentications between the video and the live study,
irrespective of the authentication length. For patterns with feedback lines (termed: PAT) ($\chi^2=0.0,p=1$), there is
strong statistical similarity. However, for patterns without lines (termed: NPAT) ($\chi^2=4.54,p=0.03$) we do
see a significant difference between the live and video study, and an even more
prominent difference for PINs ($\chi^2=37.76,p=0.00$). Statistical differences for NPAT can be accounted for by an increase in the 4-length performance for attackers in the live setting (see
Table~\ref{tab:views}), and for PINs, we consistently see performance increases
for the live setting compared to the video setting. In this case, the success
rate for PINs in the video setting is 65/208=32.0\% compared to 135/216=62.5\%
for the live setting, an increase of 1.95x; however, the video study does
provide a baseline.

We can also compare authentication types within collection method, as related to
{\bf H1-p}. Using a three-way $\chi^2$ tests with pairwise comparisons,
there are statistically significant differences between each of the success rates
for each of the authentications for both the live ($\chi^2=24.8,p=0.00$) and
video ($\chi^2=133.4,p=0.00$) settings. The residuals suggest the leading cause
of this difference is the increased difficulty of shoulder surfing PINs, for
both the video and live setting, but post-hoc, pairwise-analysis (with
Bonferroni correction) suggest the benefits of removing feedback lines in NPAT is
not consistent across studies. While there are statistical differences between
PAT and NPAT in the video study, this effect disappears in the live study with
$p=.147$ (under the correction). {\em This provides further evidence that
  removing traceback lines from pattern entry provides limited protection, and
  perhaps less than what was previously considered~\cite{vzw2015easy}.} 

Despite seeing a reduced benefit from NPAT as compared to PAT, we can {\bf
  confirm} {\bf H1-p} in the live setting. The authentication type has an
impact on shoulder surfing performance as evident in the differences in attacker
success rate for different authentication types, particularly for PINs.

\tableangles{}

\paragraph{Repeated Viewings (H1-r/H2-p)}
An important result of the video study was the finding that repeated viewings
have significant impact on attacker performance ({\bf H2-p}). By expanding our
view of Table~\ref{tab:views} to the {\em Two-Same} and {\em Two-Different}
column, we can test for similar effects resulting from repeated viewings.  As
before, we observe the most consistency in the PAT and NPAT settings for the
live and video study, and strong differences in the PIN setting. However, where
we do see significant difference the confidence interval suggest that the video
study does provide a baseline to the live setting.

We can further directly measure the impact of multiple viewings by performing
within collection method $\chi^2$ tests across viewing methods. For PAT, no
effect could be identified for multiple views in both the video and live
settings. There is an effect for NPAT in the live ($\chi^2=12.0,p<0.01$) but not
in the video setting ($\chi^2=5.1,p=0.08$).  Post-hoc analysis revealed that, for
NPAT in the live setting, having the same viewing angles twice compared to a
single viewing angle or two difference angles drives this difference ($p=0.03$,
corrected), {\em suggesting that two-different viewing conditions for NPAT is
  most advantageous to an attacker.}

The case is similar for PINs. In the live setting, a statistically significant
difference occurs for conditions of repeated views ($\chi^2=23.1,p=0.00$).
However, this was not the case for the video setting ($\chi^2=4.1,p=0.14$). {\em
  Post-hoc analysis suggests that gaining any repeated viewing, the same angle
  twice or two different, benefits the attacker significantly in the live
  setting.} The lack of significance for the video setting may be due to using
this particular subset of video data, but we conjecture that it more likely
reflects the high difficulty of shoulder surfing PINs, generally, which was
further exacerbated by the video observation setting without stereo vision.

Overall, we can {\bf confirm} {\bf H2-p} in the live setting, that repeated
viewings have an impact on performance. Where there were previous significant
differences in the video study, these persisted in the live setting, except for
NPAT. While there is consistency in viewing the same angle twice, observing the
entry from multiple angles seems to play a larger role in the live setting
compared to the video setting. However, the larger hypothesis that repeated
views impacts performance of shoulder surfing is confirmed.

\paragraph{Observation Angle (H1-r/H4-p)}
To assess the impact of observation angle, we use only single-view conditions so as
not to conflate the results with the impact of multiple observations. These results are
presented in Table~\ref{tab:angles} with pairwise comparisons between the live
and video study for different passcode lengths.

While we continue to see significant differences for PIN and a lack thereof for
PAT, we see significant improvements in the live setting for NPAT viewed from the
right angle. We conjecture that this improved attacker performance relates to
being able to stereoscopically determine touch locations that are more
challenging to see from the same angle via video simulation. However, depth of touch
events continue to be more challenging when viewed from the left angle. The
difference between the observations angles here may also explain other
statistical differences in the previously presented results for NPAT.
 
However, overall, we {\em do not} see significant differences when comparing
within a collection method and authentication when comparing left vs. right
angle. This is in conflict with prior work; however, recall that the top
observation angle was removed and the two near and far angles were reduced to a
single side angle (L or R). As the two comparable subsets are consistent, we can
{\bf confirm} that under {\bf H4-p} the live settings are well predicted by a
comparable subset of video data.

\tablecodes{}
\paragraph{Passcode Properties (H1-r/H5-p)}
In Table~\ref{tab:codes}, again using single view data, a direct comparison
between each of the passcodes used in the study is displayed, with findings from
proportionality tests between the live and video setting. We find that 
no significant differences exist for the PAT and NPAT codes, and only
three of the PIN codes show differences. These include the following PINs: 5962, 159428, and 366792 with the live setting attacker performance being significantly better in each case. The spatial
properties of these codes (see Appendix~\ref{fig:pins}) does not suggest that a
single factor played a role. Although both 5962 and 3669722 are both right
shifted PINs, there are too many other features at play to draw conclusions.

We can perform a within-collection method analysis across the passcodes using a
$\chi^2$ test, and we find that  significant differences exist for the
attacker success rate within both the live and video study, for all
authentication types. However, post-hoc analysis suggest that none of the NPAT
pairwise comparisons are significant, and only one set of PAT pairwise comparisons are
significant (743521 vs. 3157) --- 743521 was the most difficult of the patterns
to shoulder surf. For PINs in post-hoc analysis, again 159428 and 366792 have
significant comparisons, particularly with PINs 7272 and 1955, which were two of
the easiest PINs to shoulder surf in comparison to 159428 and 366792, two of the
most difficult to shoulder surf.

Finally, we can compare the impacts of length. For PAT, we do not see
significant differences between success rate for 4- vs. 6-length patterns
($\chi^2=2.9,p=0.09$), but we do for the video study
($\chi^2=12.83,p<0.001$). We find the reverse for NPAT, where there is a
significant difference in length for the live setting ($\chi^2=5.7,p=0.02$) and
not for the video study ($\chi^2=3.64,p=0.06$). Finally, we see significant
differences for PIN for both live ($\chi^2=28.9,p=0.00$) and video
($\chi^2=27.6,p=0.00$). This suggests that, yes, the length of the passcode can
have an impact, {\bf confirming} {\bf H1-r} for the {\bf H5-p} condition;
however, other properties of the passcode were not significant, but were so in
similar ways between the two studies under the subset being evaluated.

\paragraph{Hypothesis {\bf H1-r}}
Based on the results presented previously, in each case we are able to find
confirmation of each of the previous hypotheses, although, we also find that
PINs are the least consistent. This suggests that researchers should be more
skeptical of results related to PIN based authentication in the video setting. In particular, the true values may be much higher. Additionally, we find
strong evidence that the differences between PAT and NPAT may be greatly
dimensioned (although still different) in the live setting. 

\paragraph{Hypothesis {\bf H2-r}}
We can confirm {\bf H2-r} that video based recreations do provide a baseline for
live simulation. Observe that in all cases where there is significant
differences between a video and live measurement in
Tables~\ref{tab:views},~\ref{tab:angles}, and~\ref{tab:codes}, the confidence
interval suggests that the live setting has {\em higher} proportionality than
that of video setting. In essence, yes, the video study provides a baseline, but
the baseline may be much lower than one may expect, as much as 1.7x.



%% file: posthoc.tex
\subsection{Post-Hoc Participant Feedback}

One advantage of the live study is that the researchers can directly observe the
strategies of the participants and the relative difficulties encountered, as
well as via post hoc questions (the precise questions are found in the
Appendix~\ref{app:posthoc}). There is no direct comparison to the Aviv et al. prior work
here, but we believe that the strategies likely mirror those used by
participants in the video study, to some extent. 


\begin{figure}[t]
  \centering
  
  \includegraphics[width=0.45\linewidth]{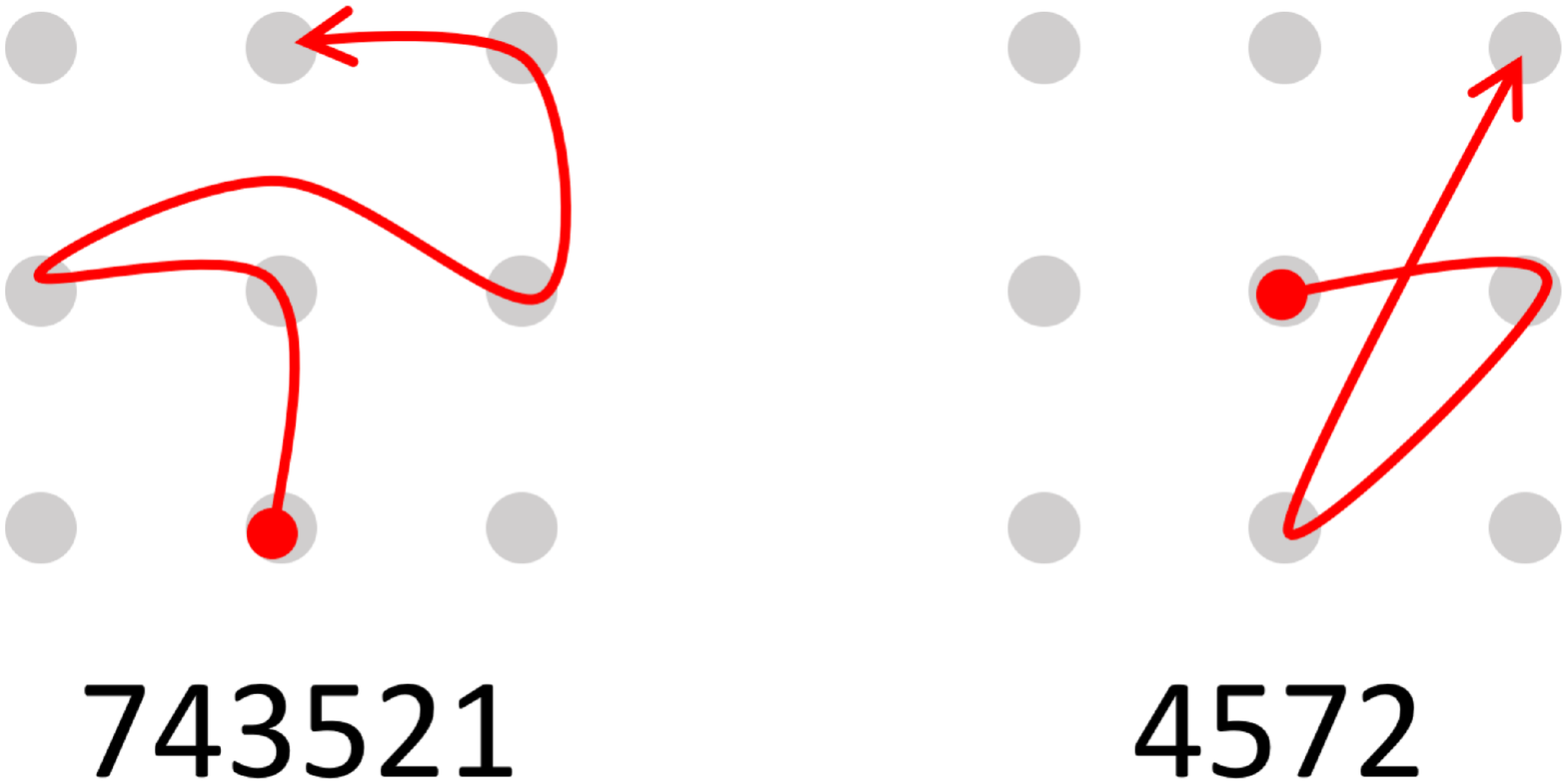}
  \hfill
  \includegraphics[width=0.45\linewidth]{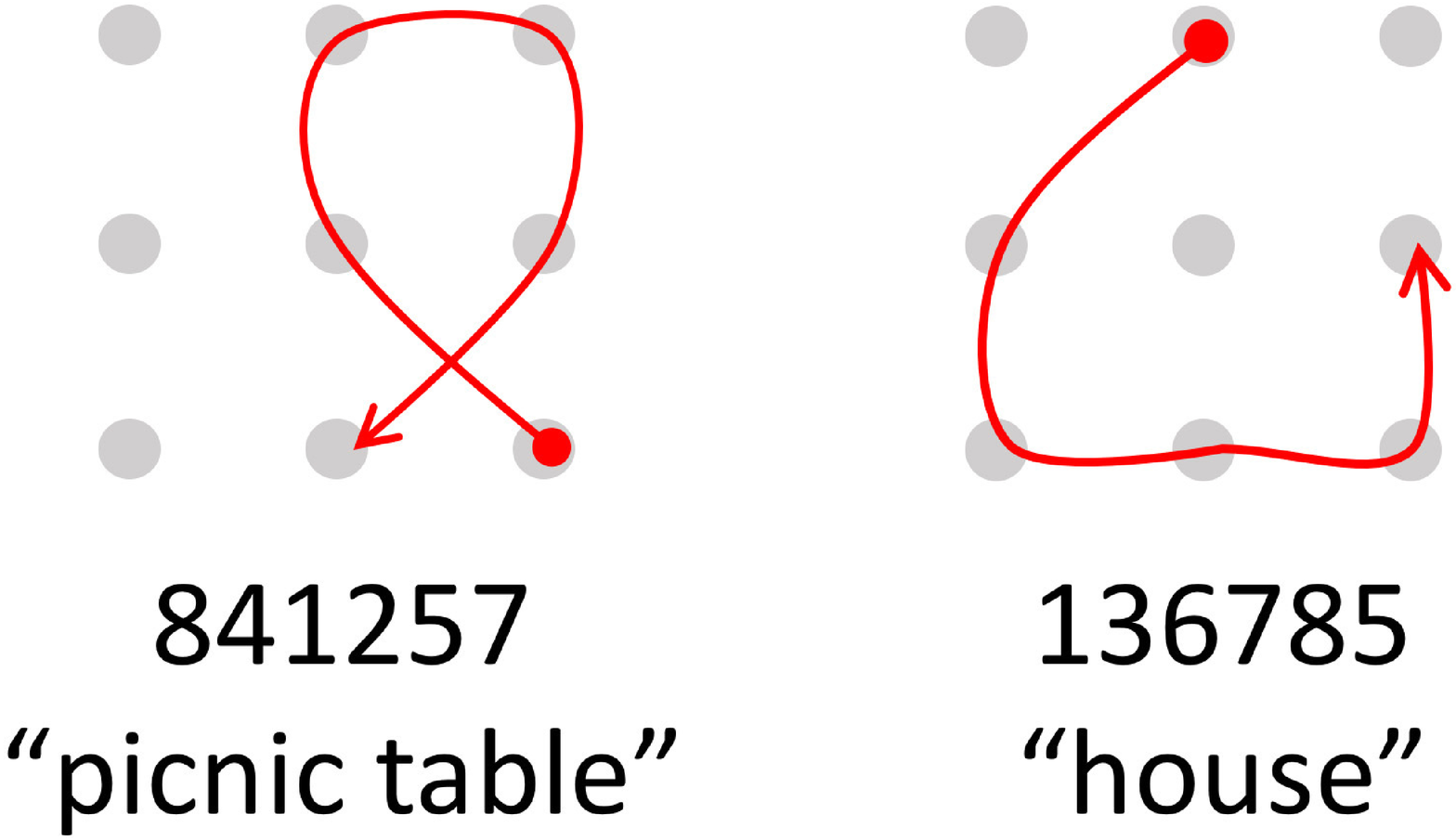}
  
  \caption{Grid patterns with crossing and knightmove (4572) features that
    challenged some observers, and patterns that were deemed more memorable by
    some observers because they offered easy symbolic associations.}
      \label{fig:knightmove_shapes}
  \label{fig:shape_associations}
\end{figure}


The most commonly reported strategy for the observation task (n=16) was simply
focusing on memorizing the passcode as it appeared and then, after it was
completely entered, writing it down immediately without delay. Only three
participants reported strategies involving writing or physically mirroring the
input gesture while it was happening.  Other participants (n=2) described
"chunking" PINs into larger numbers (e.g. "seventeen" versus "one-seven") in
their first languages (Farsi and Chinese) to make quick memorization
easier. Five participants mentioned that they watched the readout field in the
PIN conditions, while others preferred to watch only the finger gesture as it
was performed.

Participants mentioned several factors that could make PIN and grid passcodes
challenging to accurately record. These included grid pattern shapes that
crossed over themselves or contained knightmoves (n=11, e.g. 743521 and 4572,
Figure~\ref{fig:knightmove_shapes}), as well as both long physical jumps between
sequential PIN digits (n=3) and sequential digits physically close together
(n=7). Ten participants reported that viewing from the right was harder because
their view of the phone screen was partially blocked by the victim-proctor's
thumb in his right-handed grip, which is supported in the data, particularly for
NPAT results. Six participants also felt that glare from overhead lighting was
sometimes an issue.

Other passcode features and conditions were described as helpful by
observers. Four participants mentioned that it was easier to memorize shapes
that they could easily associate with a visual image, such as 136785 as a house,
or 842157 as a picnic table (Figure ~\ref{fig:shape_associations}). 

Finally, multiple observations of the same passcode were commonly deemed helpful
for confirming or piecing together sequences, although one participant stated
that it was easier to do this if both observations were made from the same
side. This is supported by the quantitative data.

%% file: takeawaysandimplications.tex
\section{Implications}

\paragraph{Importance of evaluating in appropriate settings}
Researchers often favor performing studies examining observational attacks with video-based stimuli presented to participants. While likely simpler to coordinate and easier to control compared to studies conducted in live settings, video studies can lack realism and are considered a methodological substitute only when necessary \cite{wiese2015pitfalls}. While findings from video studies can be helpful to determine attack rate, our findings suggest that researchers evaluating authentication interfaces should be aware that there is no substitute for testing in live settings, as the video baseline may greatly underestimate the threat of an attacker. The video baseline may serve as a method for a preliminary assessment.

\paragraph{Factors which should be taken into account when performing observational attack studies}
While factors such as authentication type and repeated views can impact attack rate, as evidenced through our study, other factors are worthy of further investigation.  Examples include examination of the impact of observational angle and spatial properties of passcodes and device screen sizes. While significant differences in some of these factors could not always be detected, subjective feedback gathered from participant observers suggested that these factors could make a difference to attacker success. Examining these in more detail, alongside gathering subjective data for purposes of identifying reasoning, is suggested to researchers, as these may play a greater role than once thought.

\paragraph{Care in selection of passcode}
Our results suggest that specific types and properties of passcode may be more susceptible to observational attack, as identified through the comparison with live settings.  As a result, users should be aware that removing the feedback lines from pattern unlock interfaces may not provide the security benefits that users expect.  Secondly, PINs are more susceptible to attack than previously identified by researchers performing video-based studies.  This is also supported in our qualitative feedback where participants noted that PINs with larger jumps were harder to attack, and for PAT/NPAT, those that are less ``shape like'' (e.g. resembling a house-like shape) are harder for participants.

\paragraph{Need for training}
As our findings have highlighted that observational attacks are more successful under specific conditions, security training for mobile device users can be developed to better understand the nature of observational threats, encouraging them to make better security choices. Some users may need to better understand what methods and parameters would provide resilience against high-probability multiple-view observation attacks mounted by ``insider threats'' \cite{wiese2016see}. Others might want those authentication factors tilted towards greater ease of use if they perceive less risk of observational attack. Better informing these choices could come in the form of interactive guidance/prompting when setting-up devices.  


%% file: conclusion.tex
\section{Conclusions}

In this paper, we have described a study comparing video recreations of shoulder
surfing to live simulation. We recreated a subset of the factors explored in the
video study and attempted to confirm prior findings in this setting. 
We were able to confirm many of the prior
claims regarding the video study, that authentication type, repeated viewings,
observation angle, and passcode properties can affect attacker performance. We
were also able to confirm that video study does form a baseline for the live
simulation; however, this baseline may be much less than desired, as much as
1.9x difference. From these findings we suggest, for researchers conducting
shoulder surfing studies with video components, that data can form a baseline
and be representative, in many situations, of what would occur in a live
simulation. However, when possible, those results should be compared to a live simulation to get a fuller picture of the data and results.


%% file: appendix_FMW.tex
\section{Survey Material}
\subsection{Ante Hoc Demographic Questionnaire}
\label{app:antehoc}
\begin{enumerate}
\item What is your age? (18-24, 25-34, 35-44, 45-54, +65, NA)?
\item What is your identified gender?
\item Do you have any physical conditions that might prevent you from observing authentication gestures performed on a mobile phone?
\item Do you use a smartphone currently? If so, what is its operating system?
\item Why did you select that phone and OS?
\item If you currently use an authentication method to lock your phone, what is the method (i.e. PIN, TouchID, grid, etc.), and why did you select it?
\item What types of mobile phone authentication have you used?  (i.e. PIN, grid pattern, password, fingerprint, face, voice, other)
\item Without telling me your current passcode, how do you select the passcodes you use?
\item How concerned are you with keeping your phone secure (1, not at all concerned, to 5, highly concerned)?
\item What experiences can you recall involving people either trying to steal or use your phone without permission?
\item What experiences can you recall involving people trying to observe your passcodes without permission?
\item How concerned are you with the threat of someone watching you authenticate and collecting your passcodes (1, not at all concerned, to 5, highly concerned)?
\item If you had any of these experiences, how did it affect your behavior?
\item Have any other experiences or concerns affected your authentication?
\end{enumerate}

\subsection{Post Hoc Participant Strategies Questionnaire Questions}
\label{app:posthoc}
\begin{enumerate}
\item What strategies did you employ to collect the passcodes?
\item Do you have any ideas for additional strategies?
\item How challenging was it to collect PIN passcodes (1, not at all challenging, to 5, very challenging)?
\item How challenging was it to collect grid passcodes (1, not at all challenging, to 5, very challenging)?
\item What features of the passcodes made it easier or more difficult to collect the passcodes you saw?
\item How did the number of views you were given make a difference?
\item How did which side you stood on make any difference?
\end{enumerate}

\subsection{Observation Forms}
\label{fig:obvforms}
\begin{center}
\fbox{\includegraphics[width=0.8\linewidth]{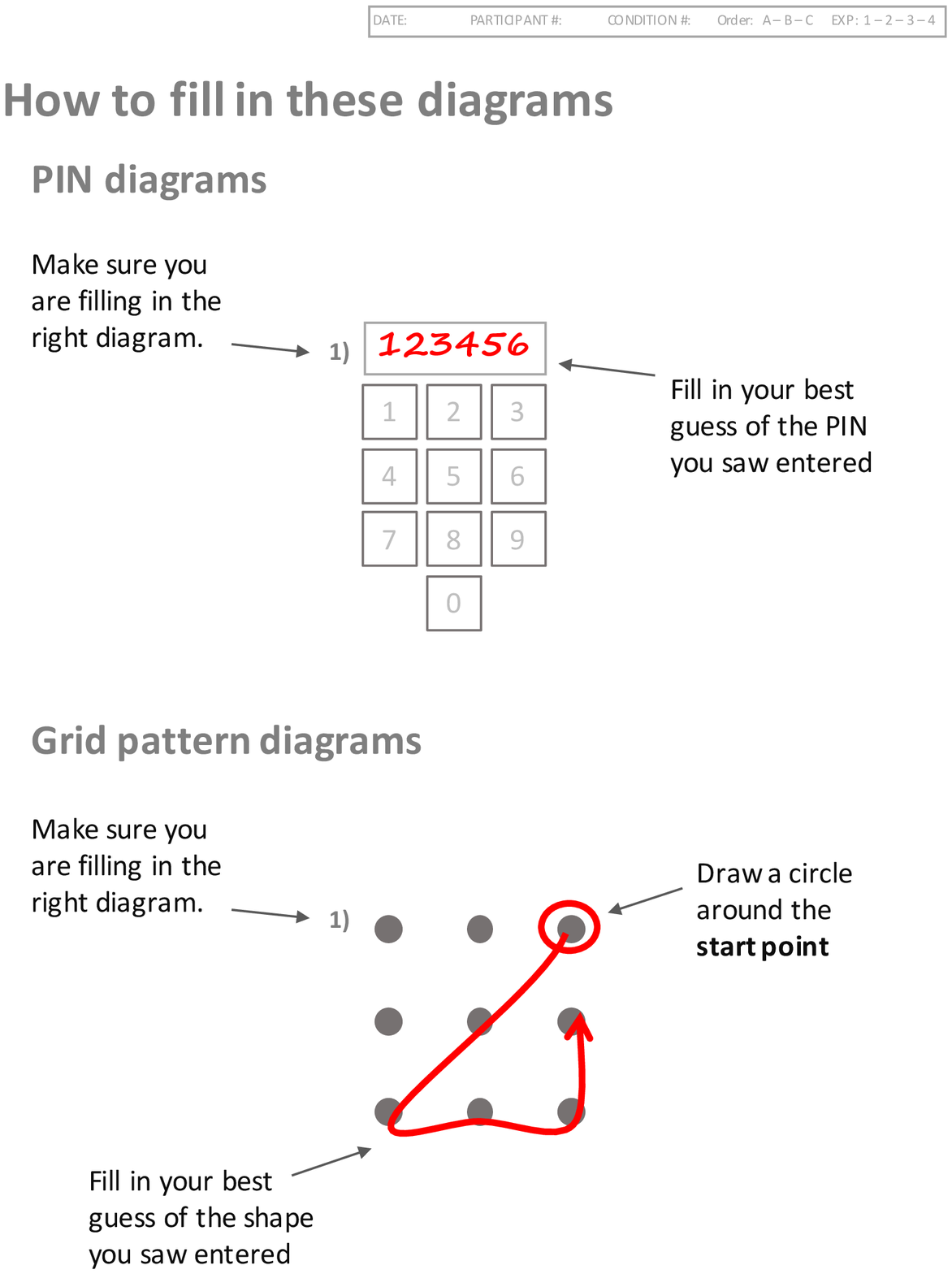}}

\fbox{\includegraphics[width=0.8\linewidth]{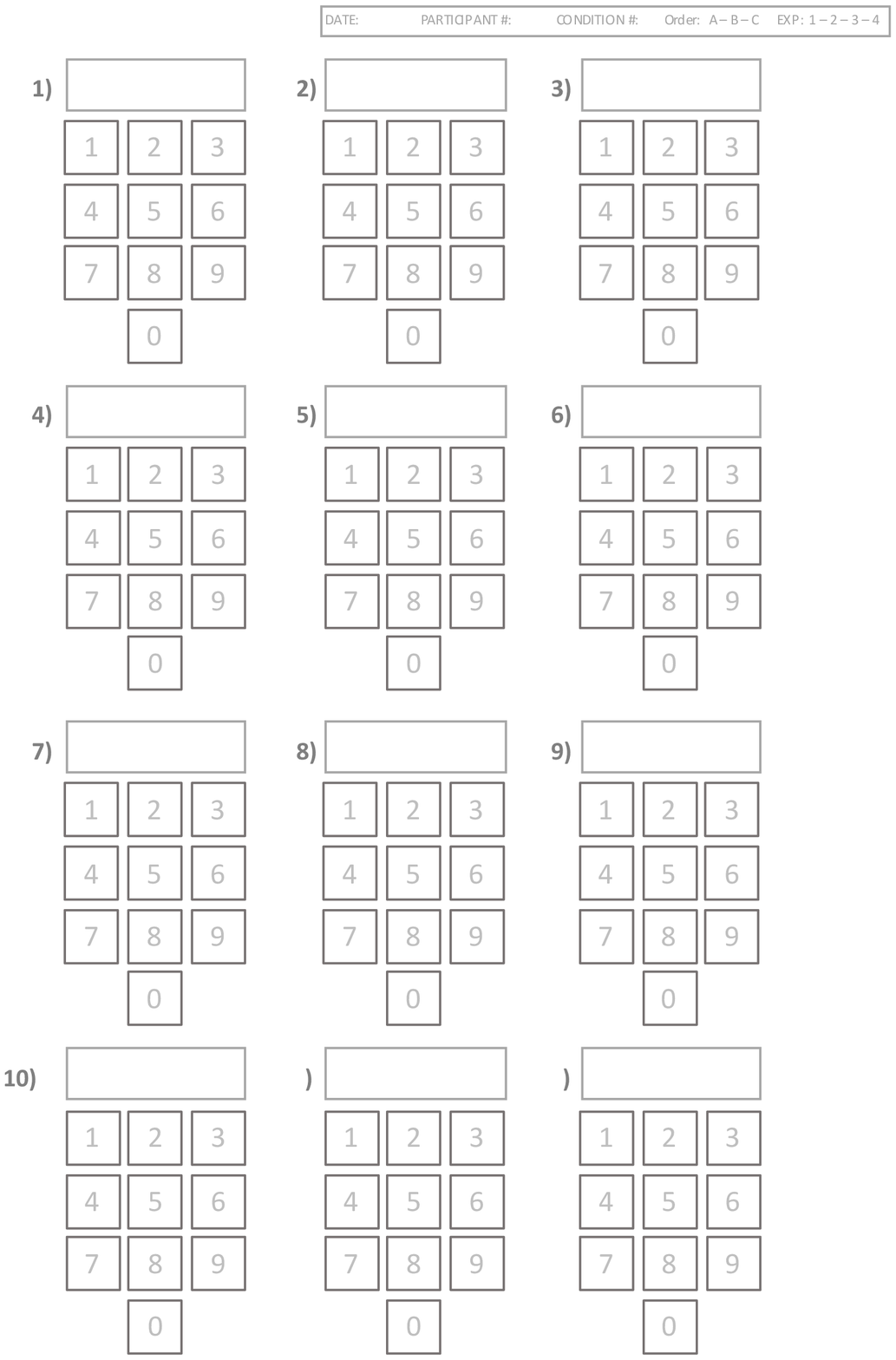}}

\fbox{\includegraphics[width=0.8\linewidth]{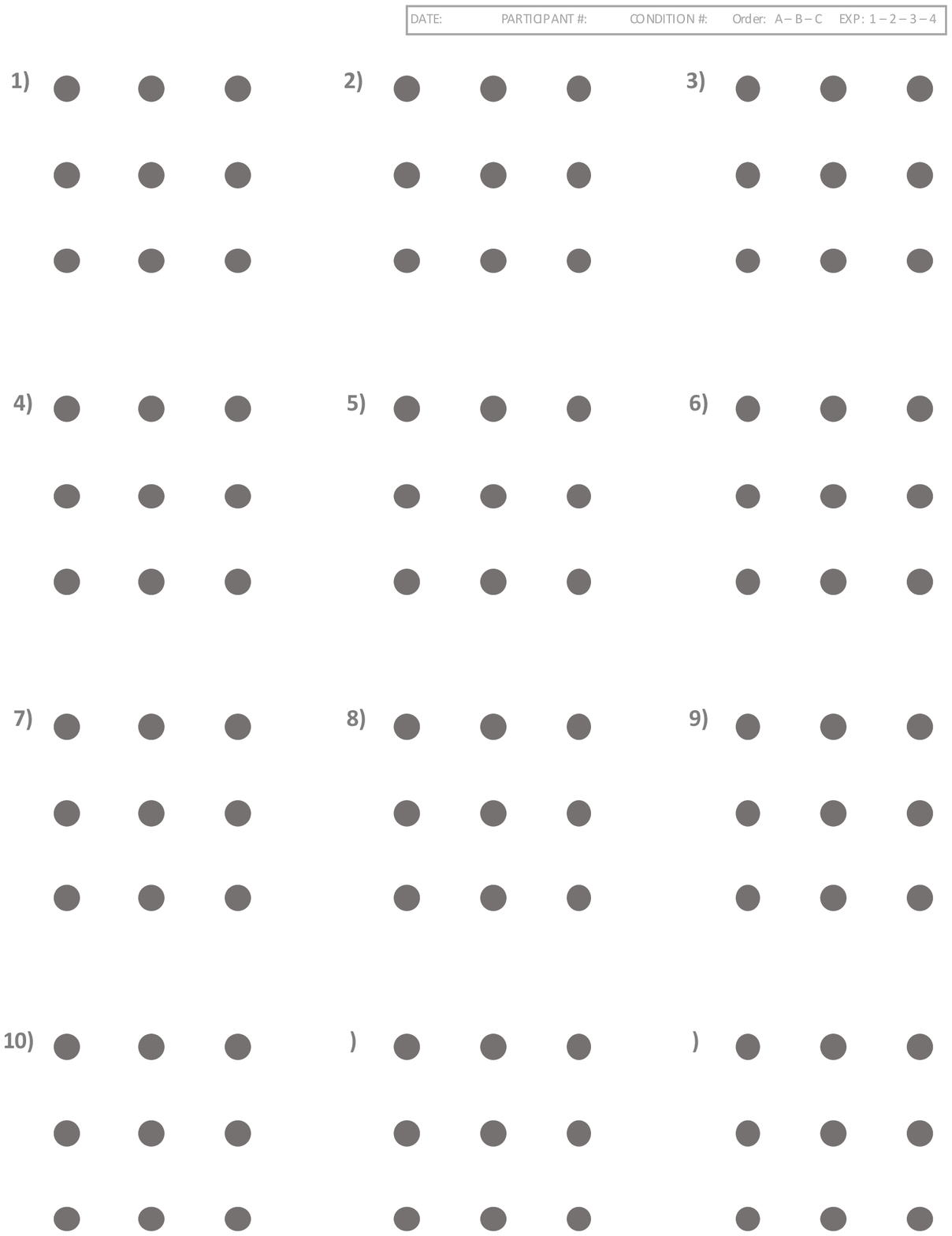}}
\end{center}

\subsection{Guide/Script for Administering Study}
\label{app:studyguide}
\begin{enumerate}
\item Verify current participant number, exp (1-4, order (a-c). Record this.
\item Introduction - "Welcome, thanks for participating. Our study deals with the security of different types of passcodes for mobile phones. Your help today will be pretty straightforward. We will record some basic demographic information about you. You will observe someone entering in passcodes from a few feet away, and write down your best guess of what you have seen. We will go over the steps required to make sure you are comfortable and understand your role. We will record the session to verify the results. All data collected will be anonymized for publication. Your part in the study should take about 20-30 minutes."
\item Payment - "The study pays 5 USD."
\item Observer disclosure - "Are there any issues, such as corrective glasses or contacts, which might interfere with performing the role I have described?"
\item IRB Introduction - "This study has been reviewed by the University's review board, the IRB, and approved as safe and ethical. Here is a copy of that form that describes the study that you can read. Please ask any questions you may have, and sign the form if you would like to participate."
\item Demographic questionnaire	- "Please fill out the demographic questionnaire."
\item Training: Overview - "Here is the process your role as an observer. You will stand to the left or right behind our researcher, who will sit in the same position entering in passcodes.  We will specify where to stand for each attempt. You will watch each attempt, and then draw on the form we provide your best guess of the passcode you just saw being entered on the phone. Passcodes may vary in length. We will repeat this ten times."
\item Training: Filling out the form diagrams - "Look at the form we have provided. It has blank PIN and grid pattern diagrams for you to enter your guesses. For the grid patterns passcodes, draw the shape you saw entered, and circle the starting point of the shape. For PIN passcodes, write out the sequence, like "1234", and draw the shape you saw entered."
[Demonstrate drawing a diagram, then allow the participant to practice drawing 2-3 times, based on a practice code for their prescribed passcode condition (PIN, grid, no-line grid) that you show them slowly, up close, on a phone. Confirm for grid shapes that they are circling the starting point. Correct any issues that appear, and repeat until ready.]
\item Training: Taking position for each attempt - "We will call out a position, LEFT or RIGHT, for you to stand in for each attempt. Move to the corresponding marker on the floor, figure out which diagram you are going to fill in, and when set to begin, say "Ready." Sometimes, we may also call out MULTIPLE if you are allowed to view the passcode being entered twice before making your final guess."
"Any questions?"
\item Post hoc questionnaire - [Conduct post hoc interview]
\end{enumerate}




%% file: pattern_viz.tex

\section{Visualization of Authentication}
Below are the set of patterns and PINs as visualized in prior
work~\cite{aviv2017shoulder}, as well as the description of the visual
properties.
\vspace{50 mm}

\subsection{Patterns}
\label{fig:patterns}
\begin{center}
\begin{tabular}{c c}
\fbox{\includegraphics[width=0.1\linewidth]{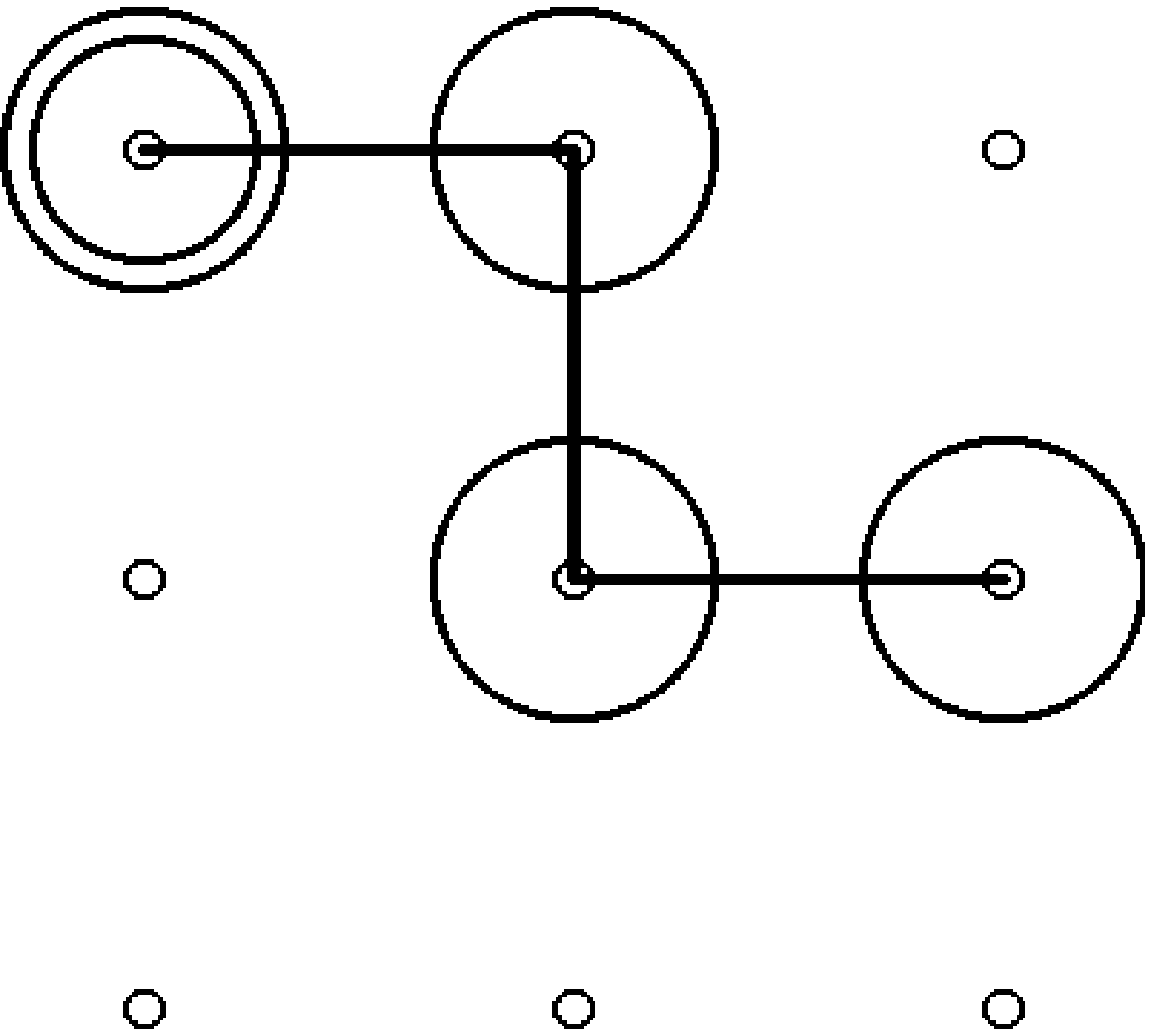}} &
\fbox{\includegraphics[width=0.1\linewidth]{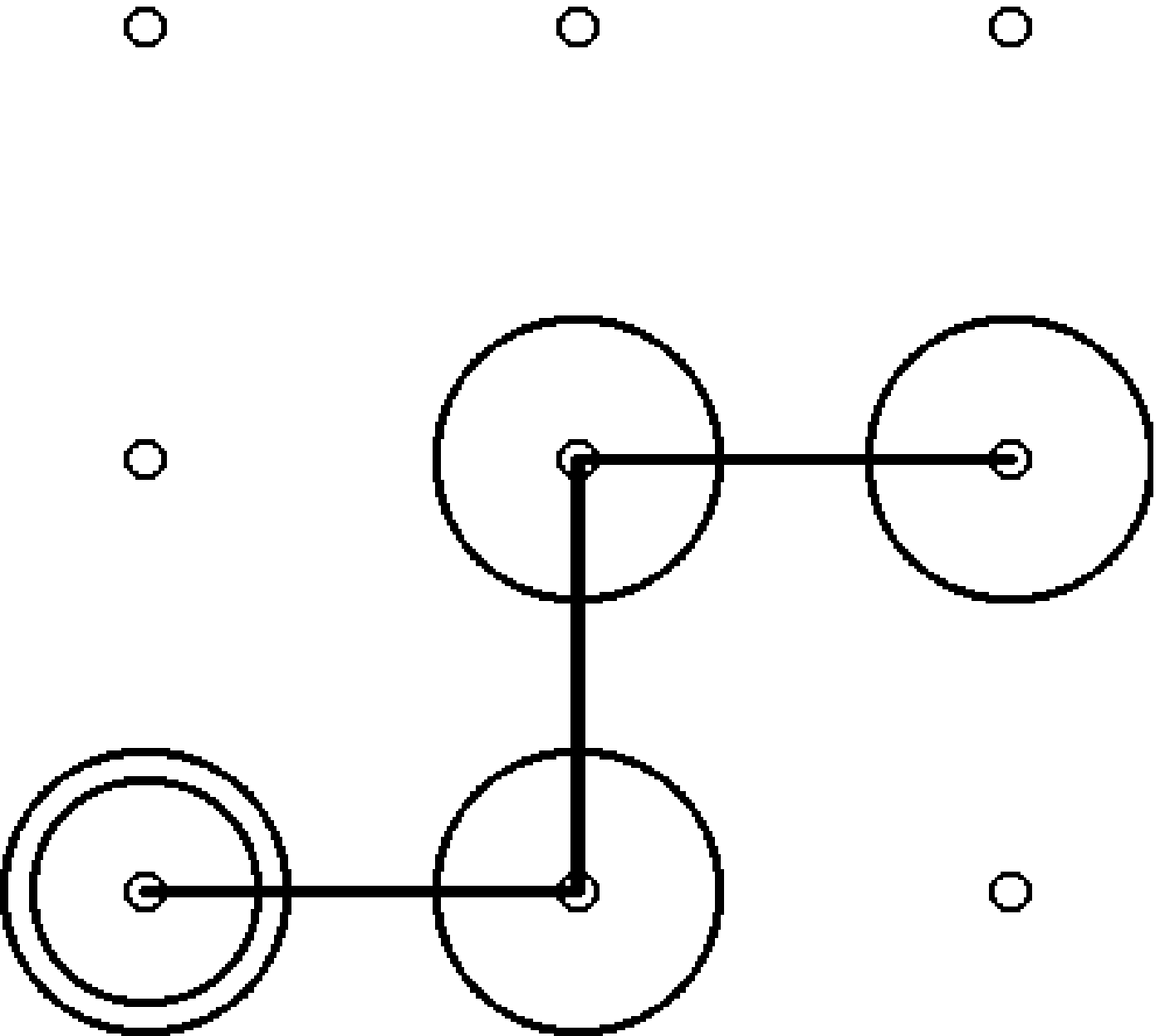}} \\
0145 & 6745 \\
up & down\\\\
\end{tabular}

\begin{tabular}{c c}
\fbox{\includegraphics[width=0.1\linewidth]{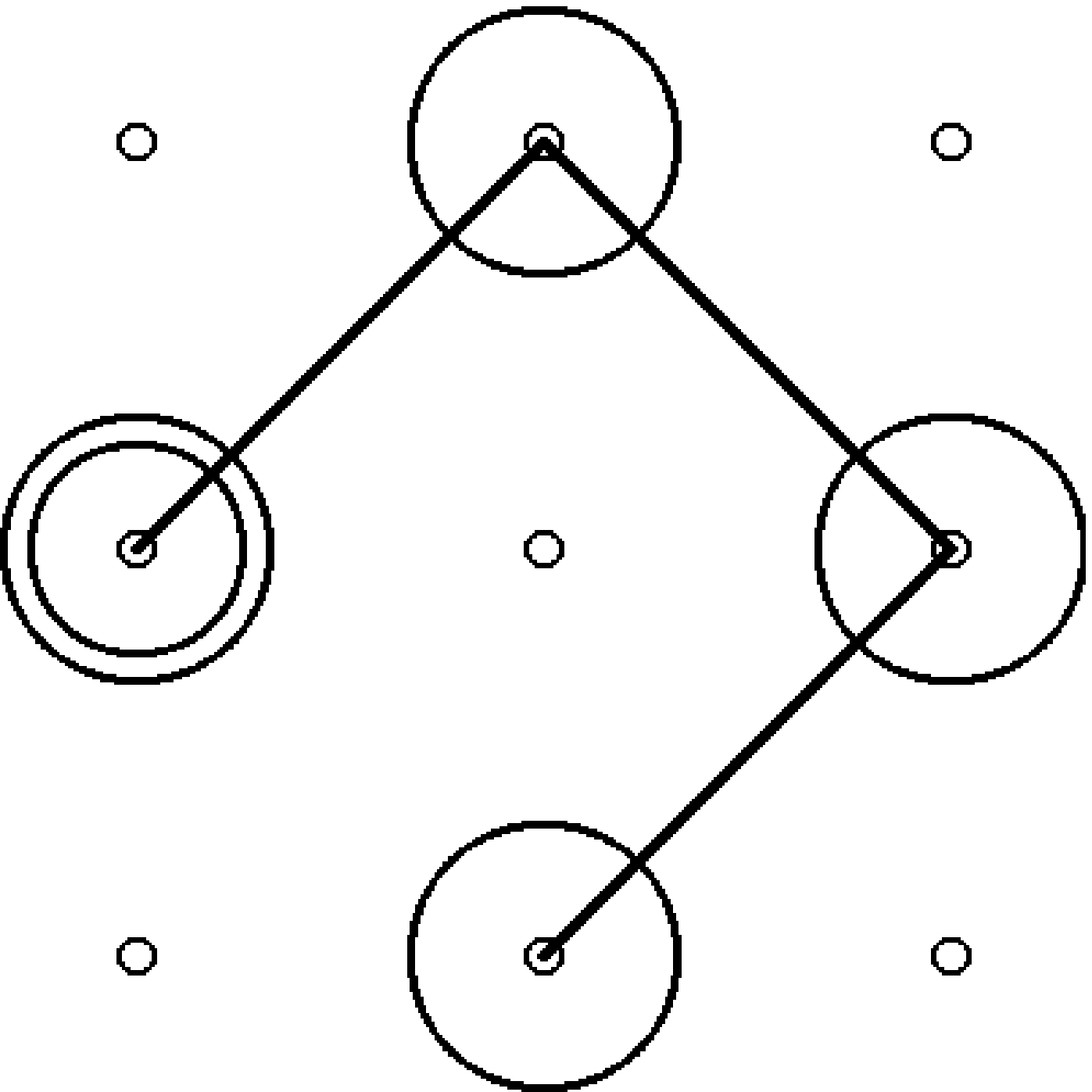}} &
\fbox{\includegraphics[width=0.1\linewidth]{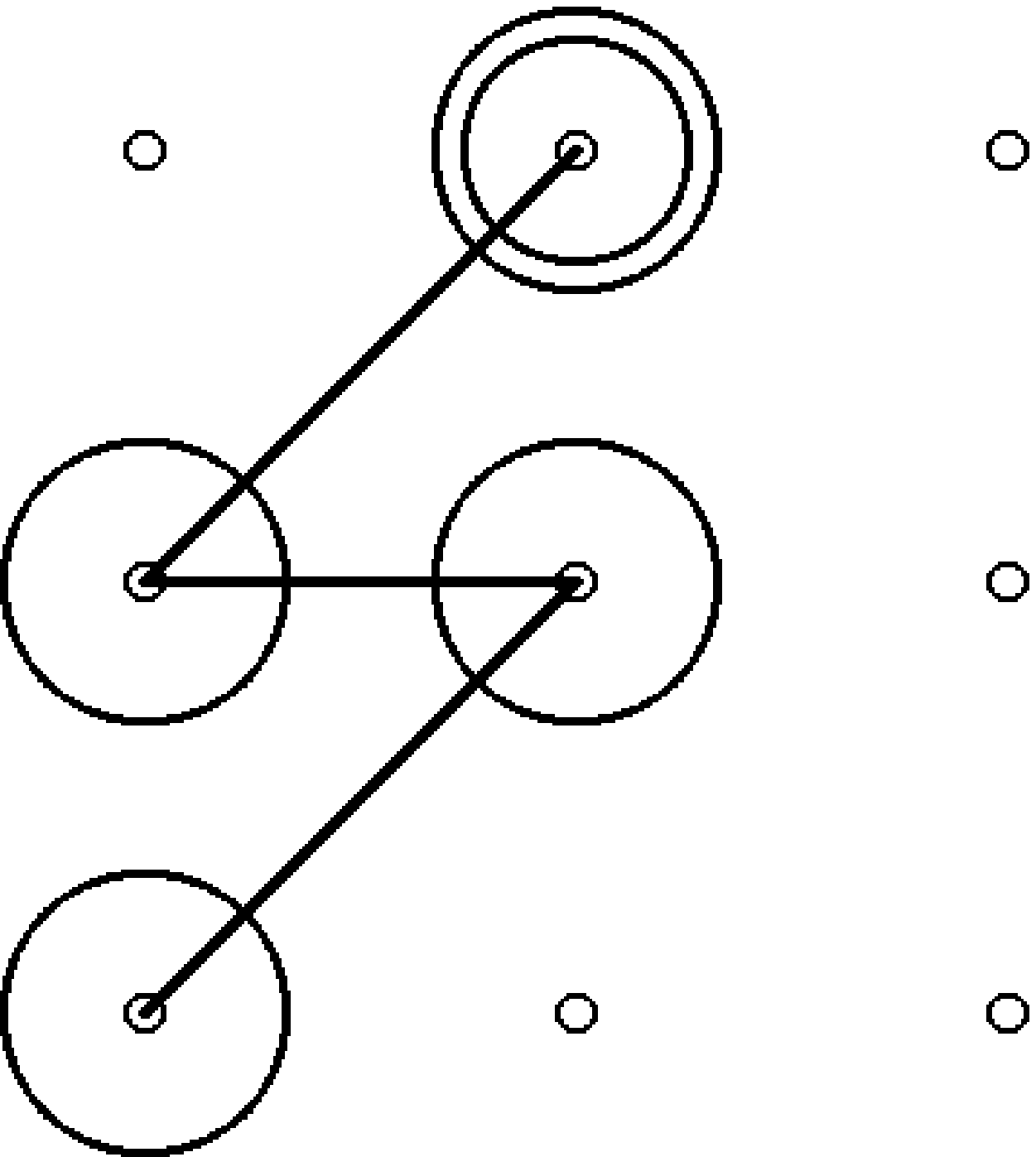}} \\
3157 &  1346 \\
 neutral & left \\\\
\end{tabular}

\begin{tabular}{c c}
\fbox{\includegraphics[width=0.1\linewidth]{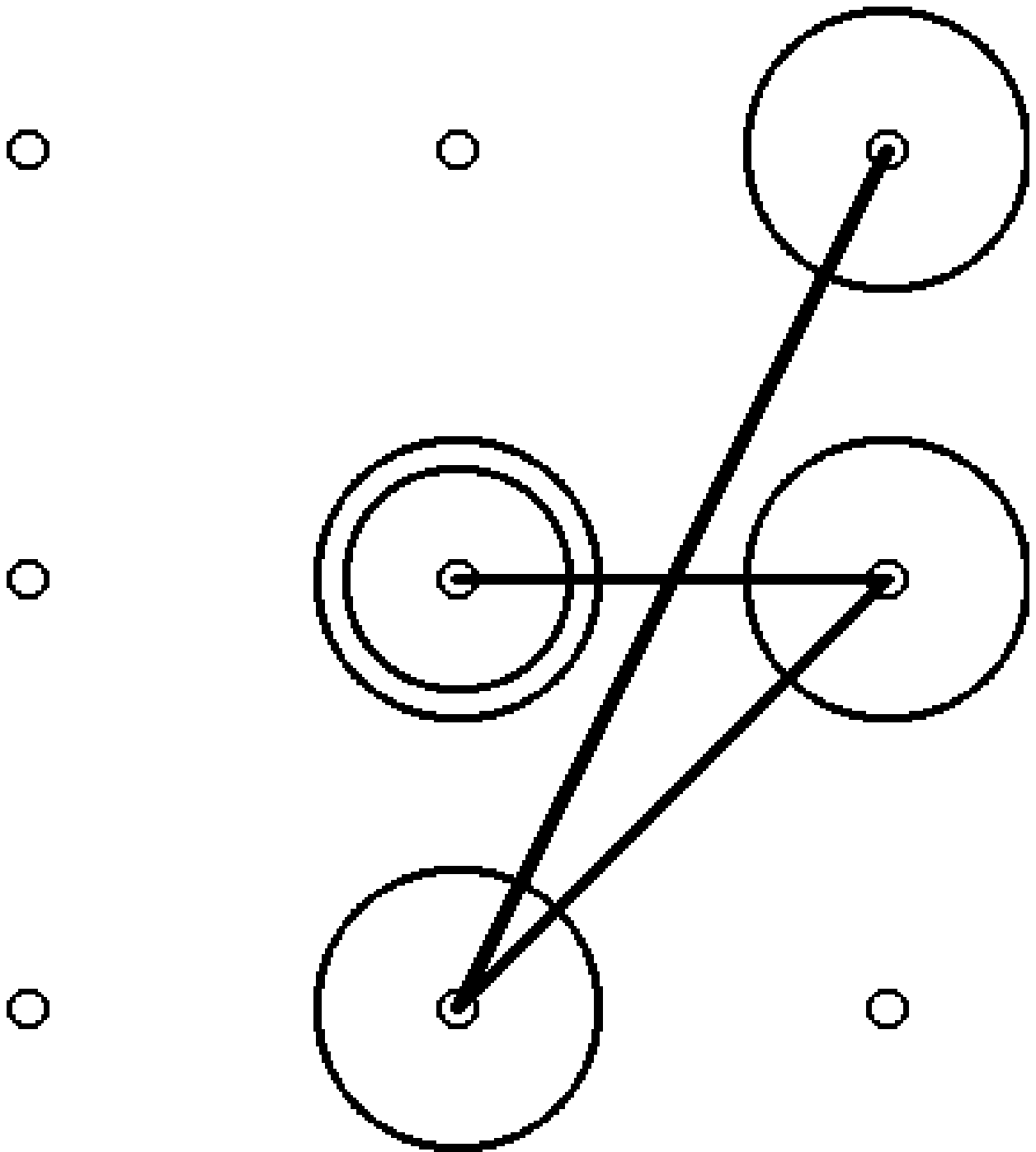}} & \fbox{\includegraphics[width=0.1\linewidth]{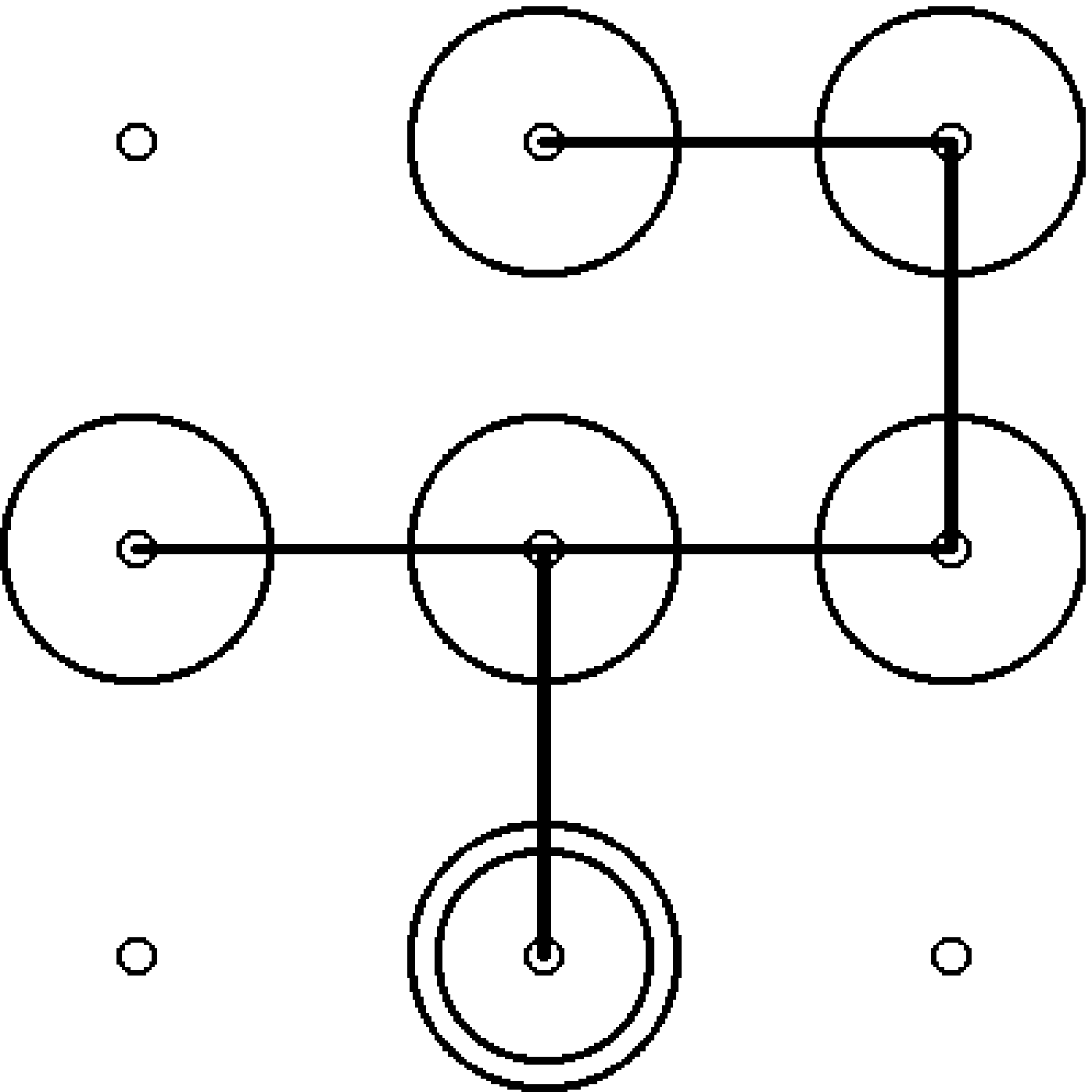}} \\
4572  & 743521  \\
right/cross & up/non-adj \\ \\
\end{tabular}

\begin{tabular}{c c}
\fbox{\includegraphics[width=0.1\linewidth]{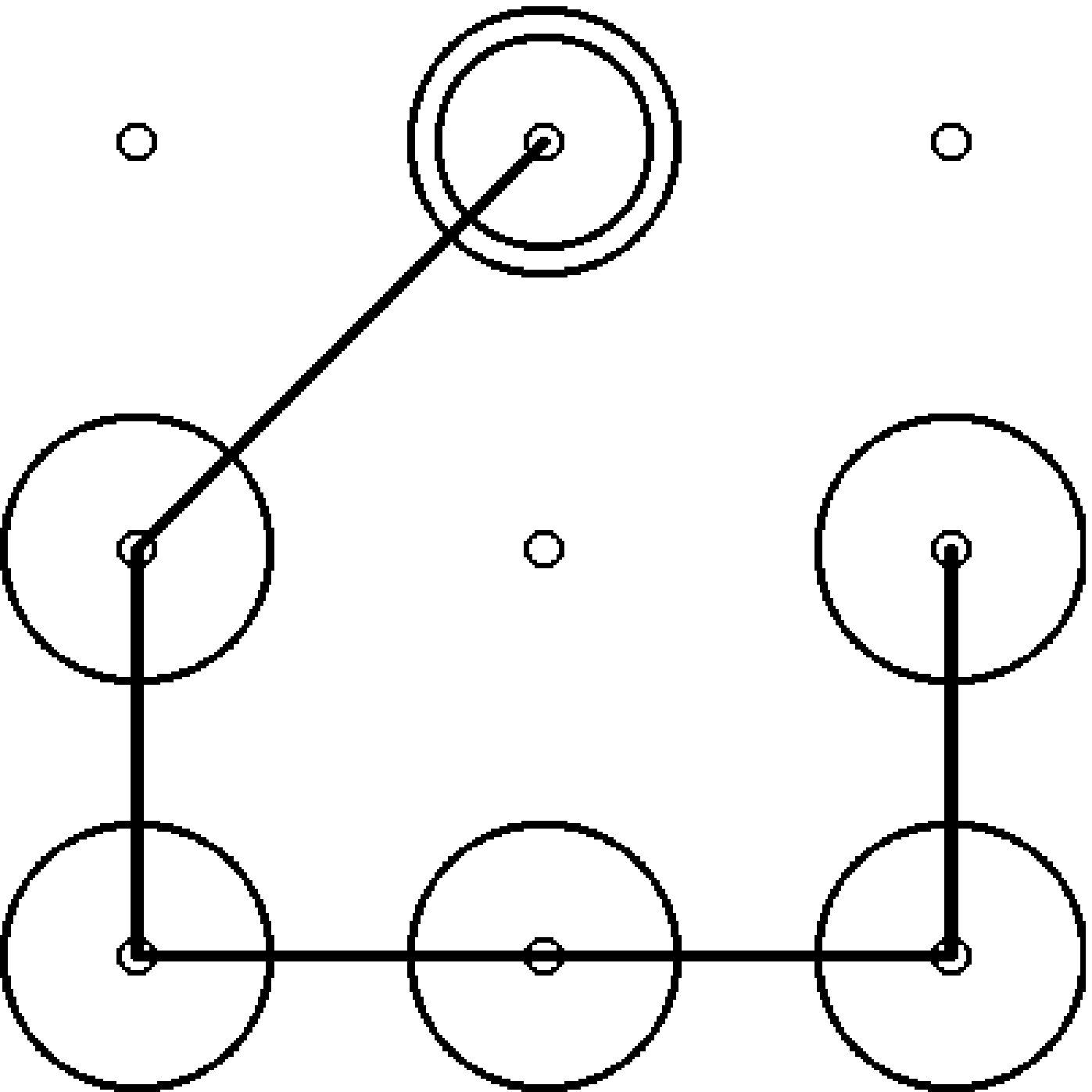}} &
\fbox{\includegraphics[width=0.1\linewidth]{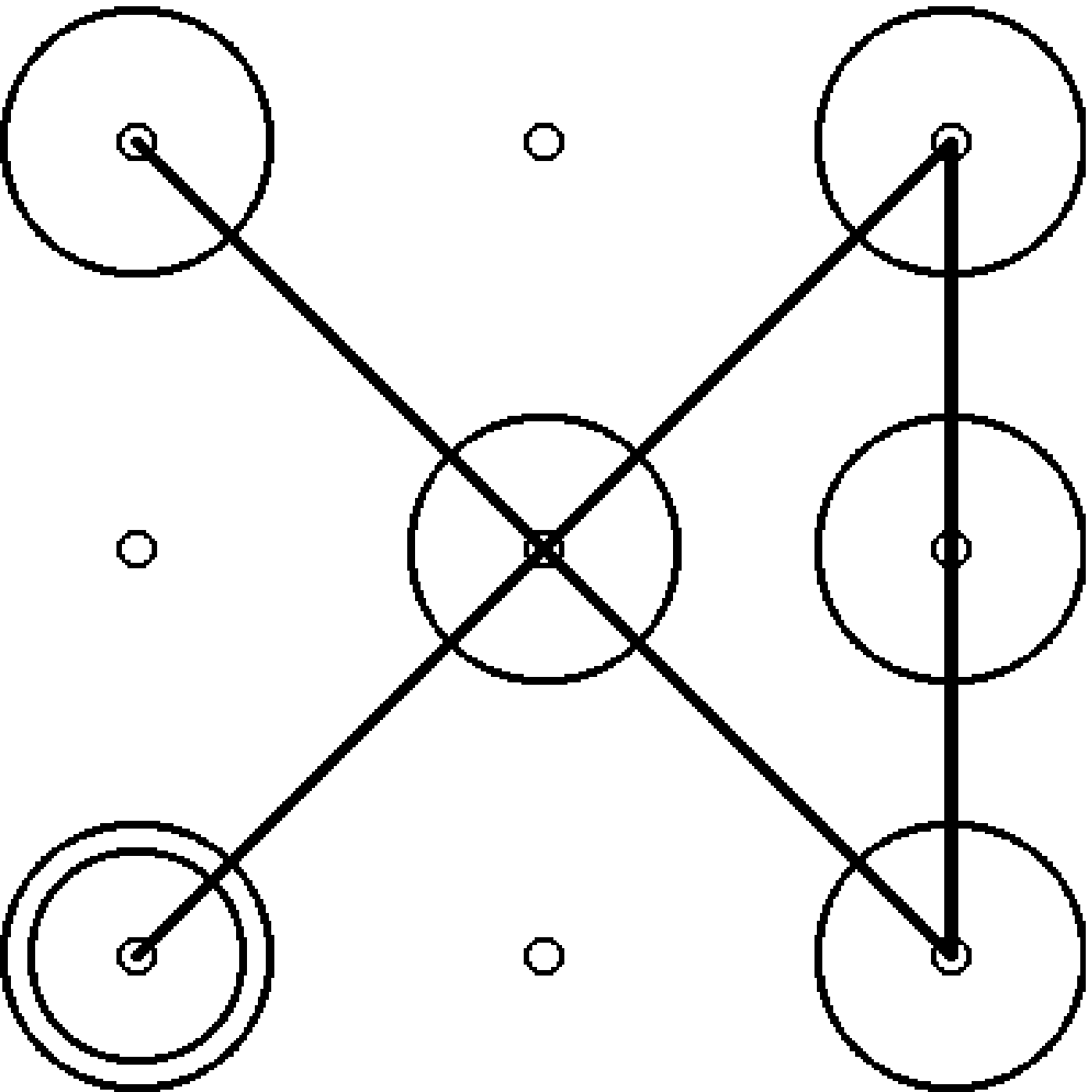}} \\
 136785 &  642580\\
 down & neutral/cross \\ \\
\end{tabular}

\begin{tabular}{c c}
\fbox{\includegraphics[width=0.1\linewidth]{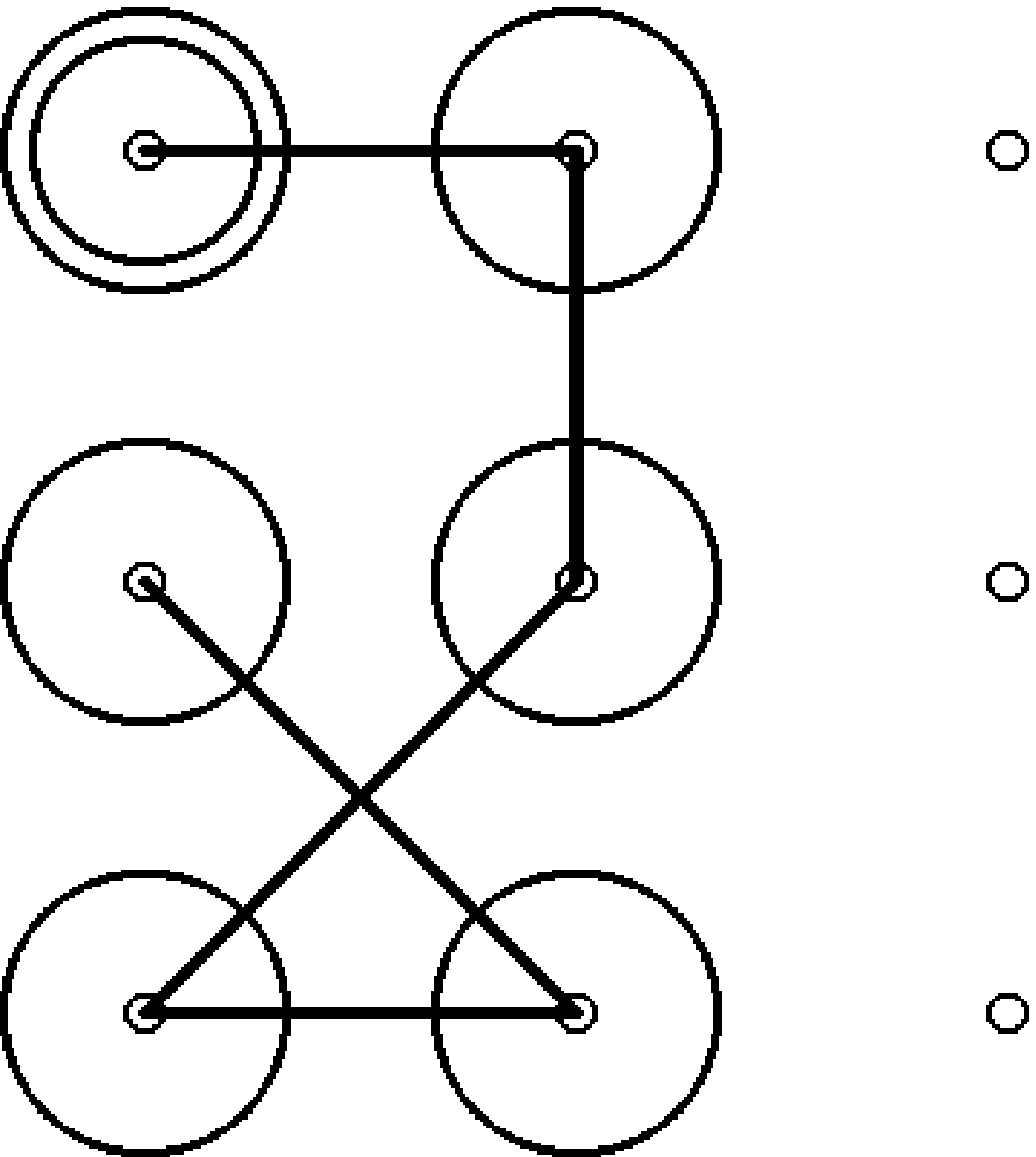}} &
\fbox{\includegraphics[width=0.1\linewidth]{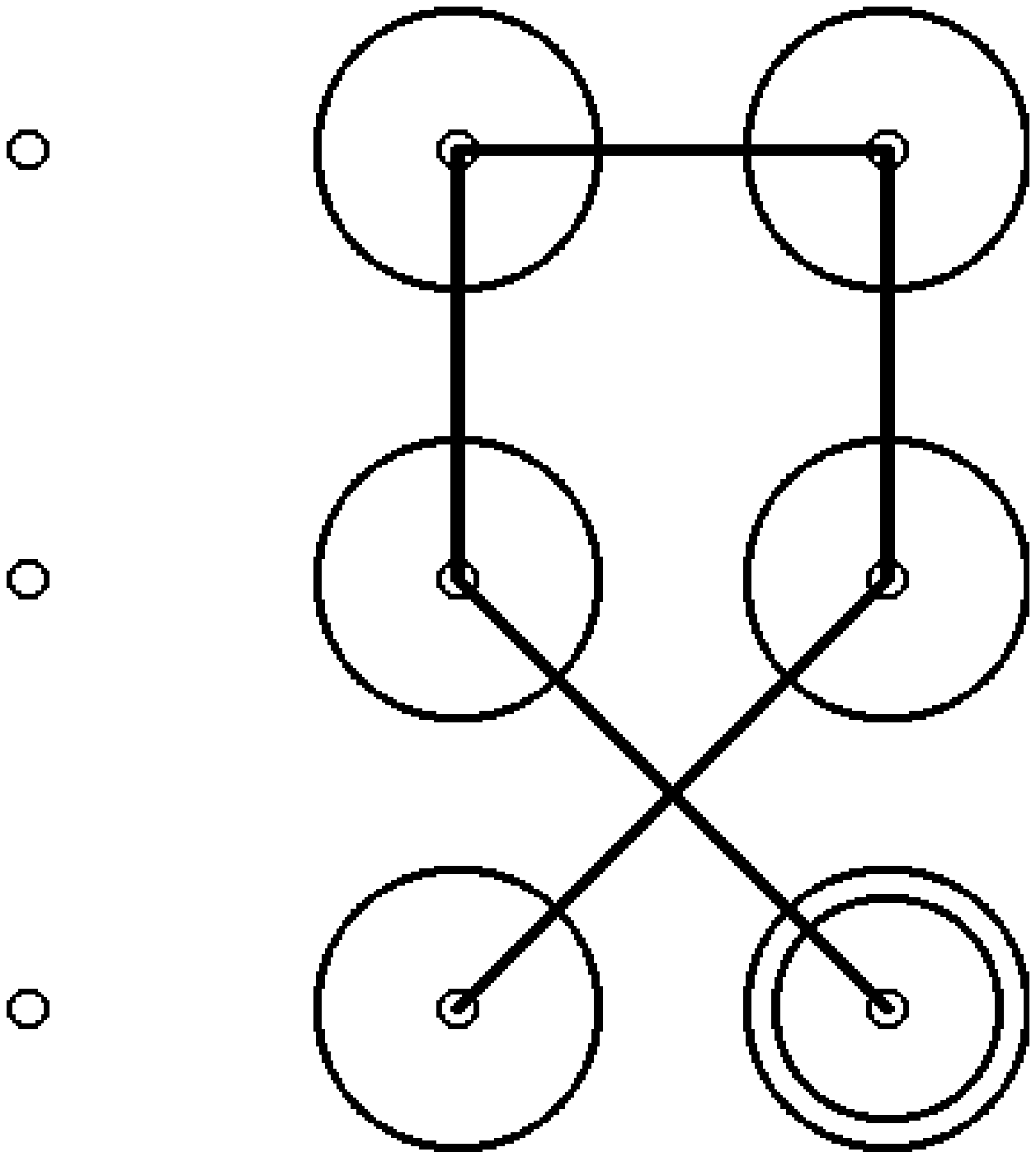}} \\
 014673  & 841257 \\
left & right/kmove/cross \\
\end{tabular}
\end{center}

\subsection{PINs}
\label{fig:pins}

Note that filled circle is the start point, multiple circles on a number indicate multiple touches. 

\bigskip

\begin{center}
  \begin{tabular}{c c}
    \fbox{\includegraphics[width=0.1\linewidth]{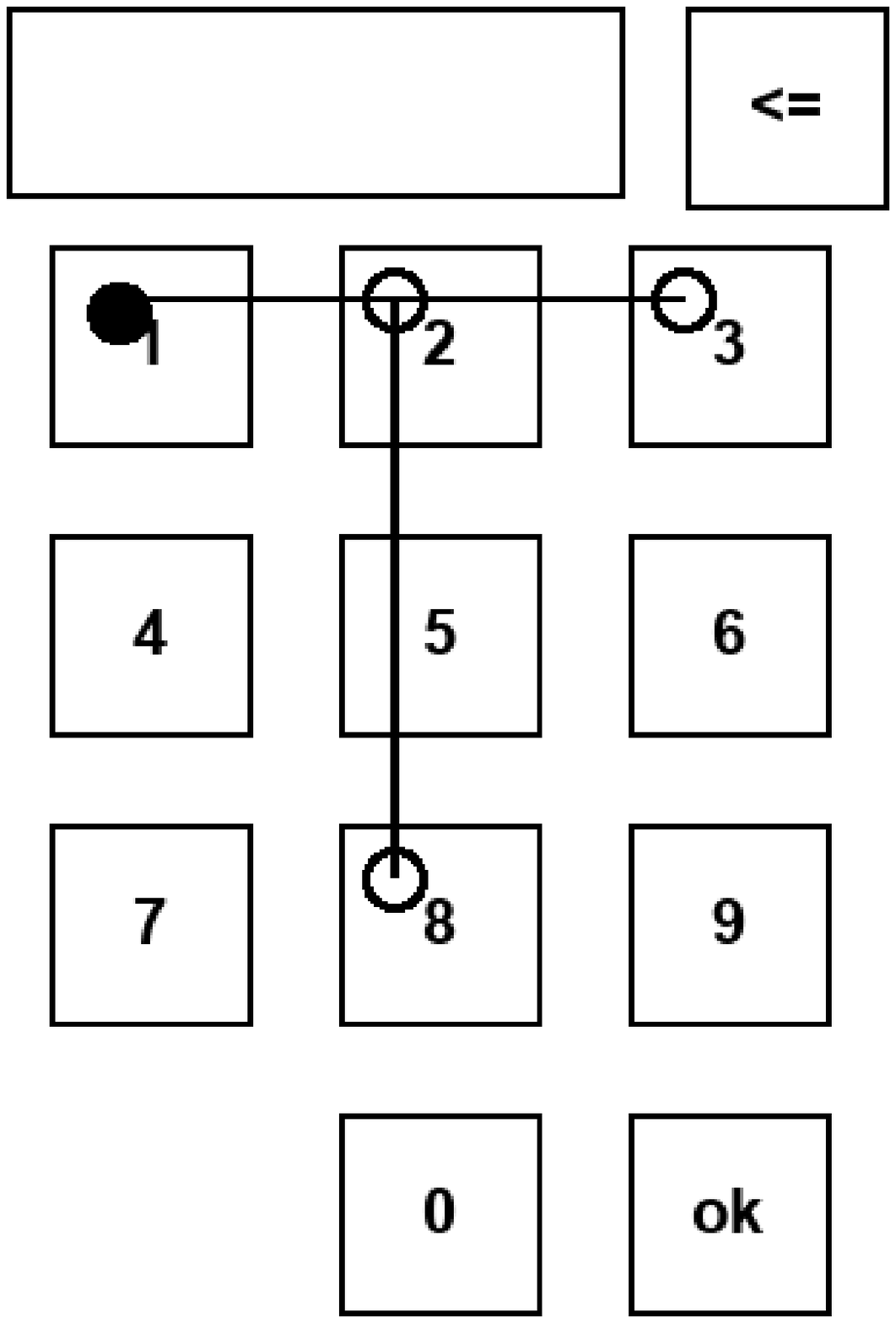}}&
    \fbox{\includegraphics[width=0.1\linewidth]{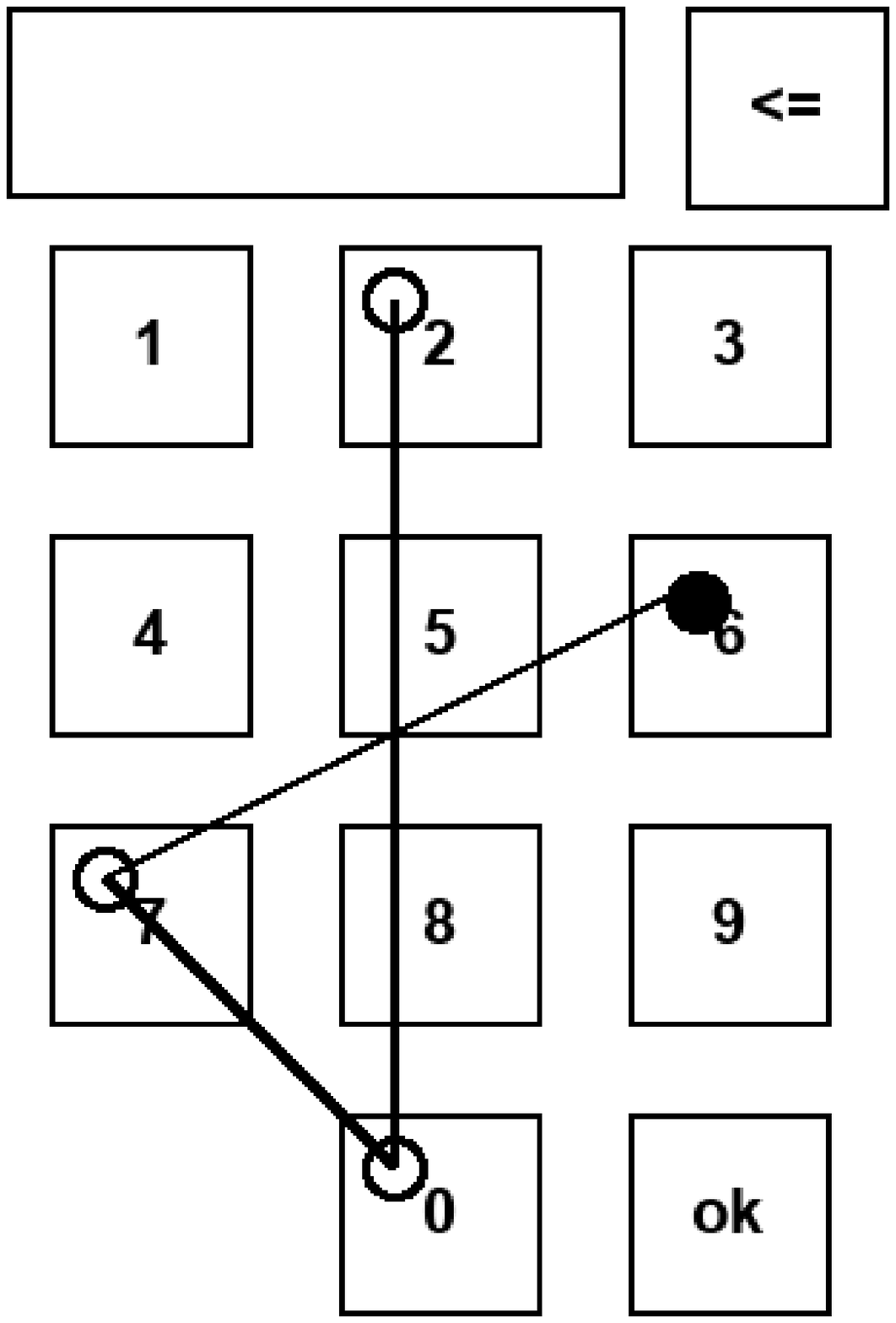}}\\
1328 & 6702 \\
up/non-adj   & down/kmove/cross\\\\
  \end{tabular}

  \begin{tabular}{c c}
    \fbox{\includegraphics[width=0.1\linewidth]{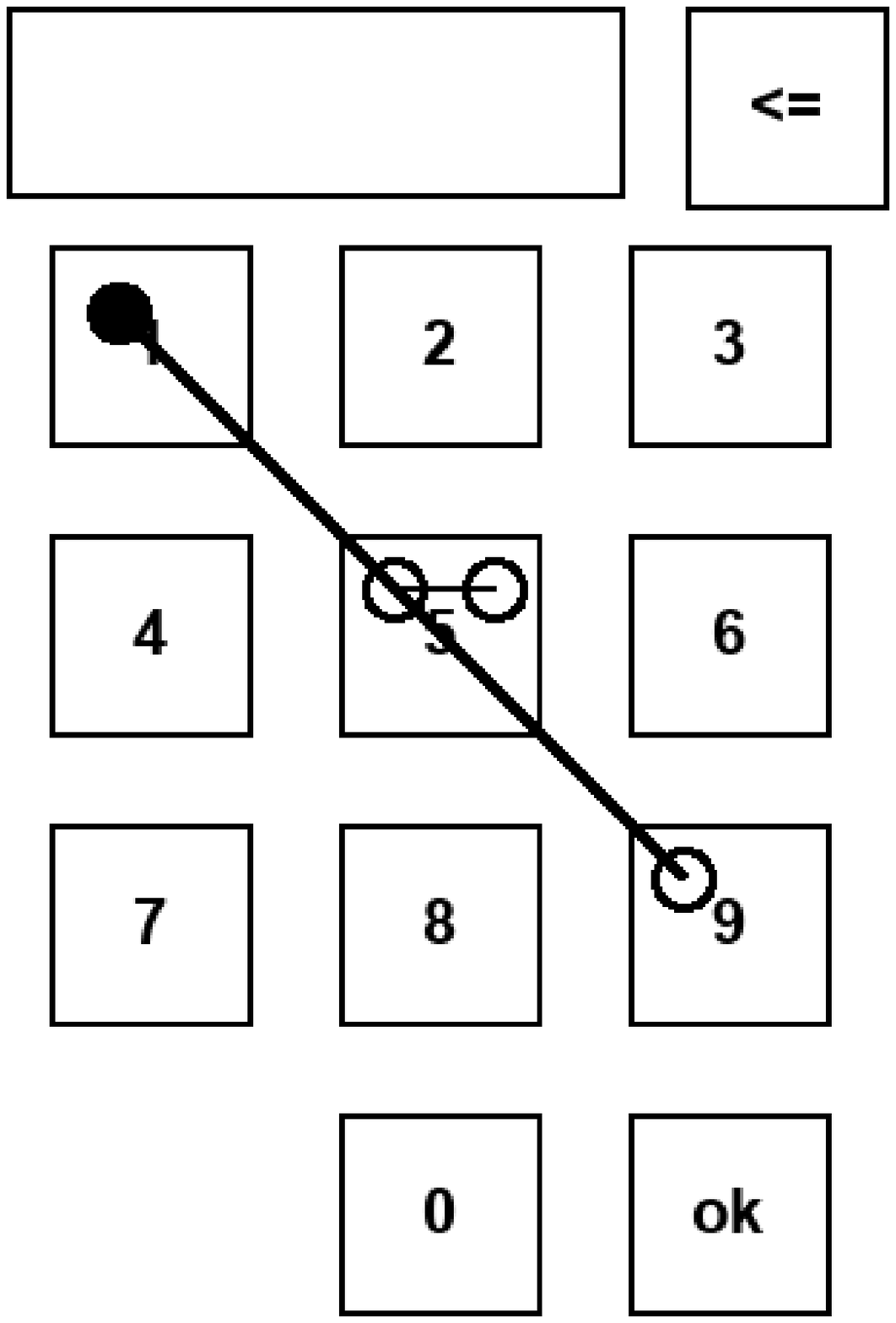}}&
    \fbox{\includegraphics[width=0.1\linewidth]{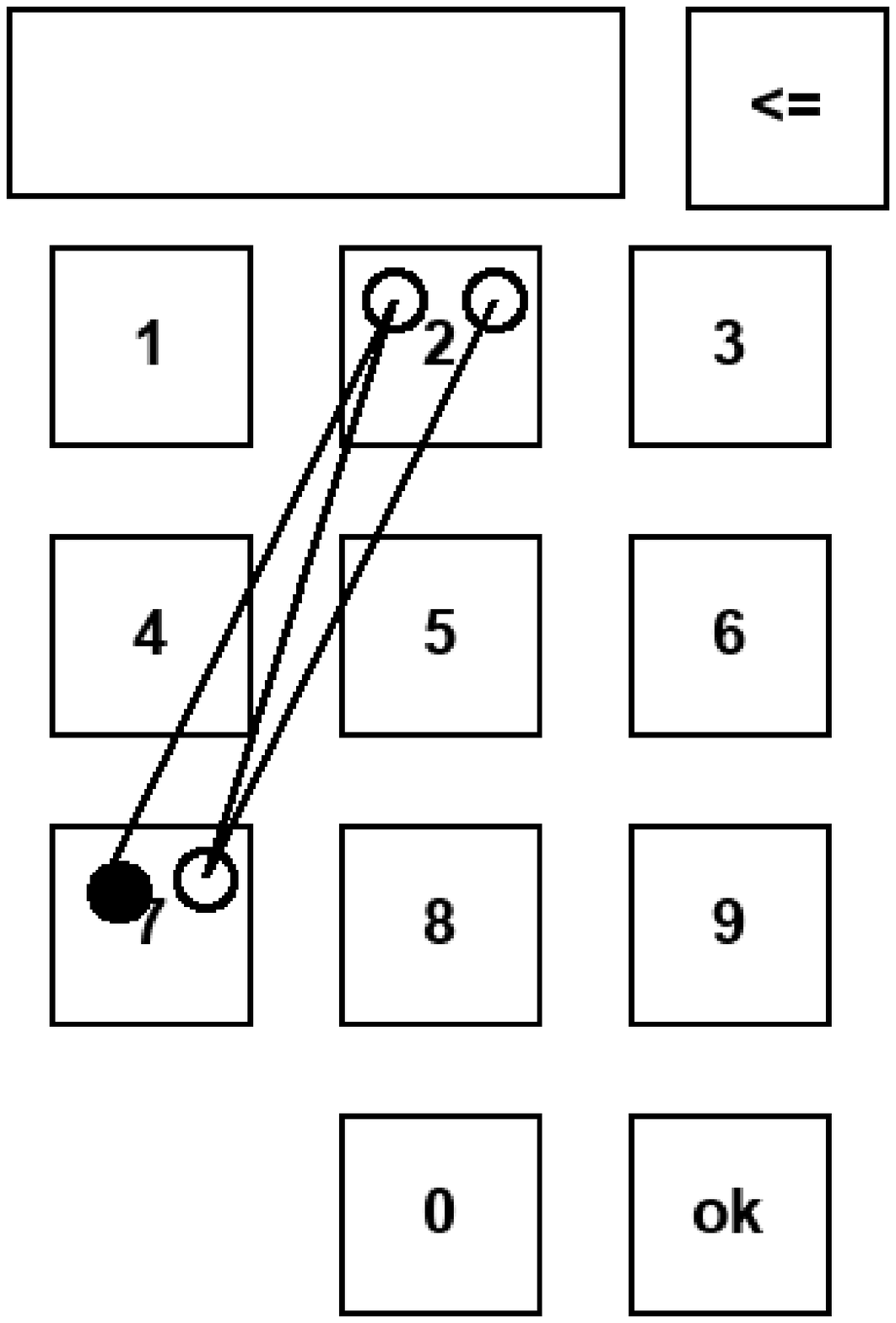}}\\
 1955    & 7272\\
 neutral/non-adj/repeats & left/kmoves/repeats\\\\
  \end{tabular}

  \begin{tabular}{c c}
    \fbox{\includegraphics[width=0.1\linewidth]{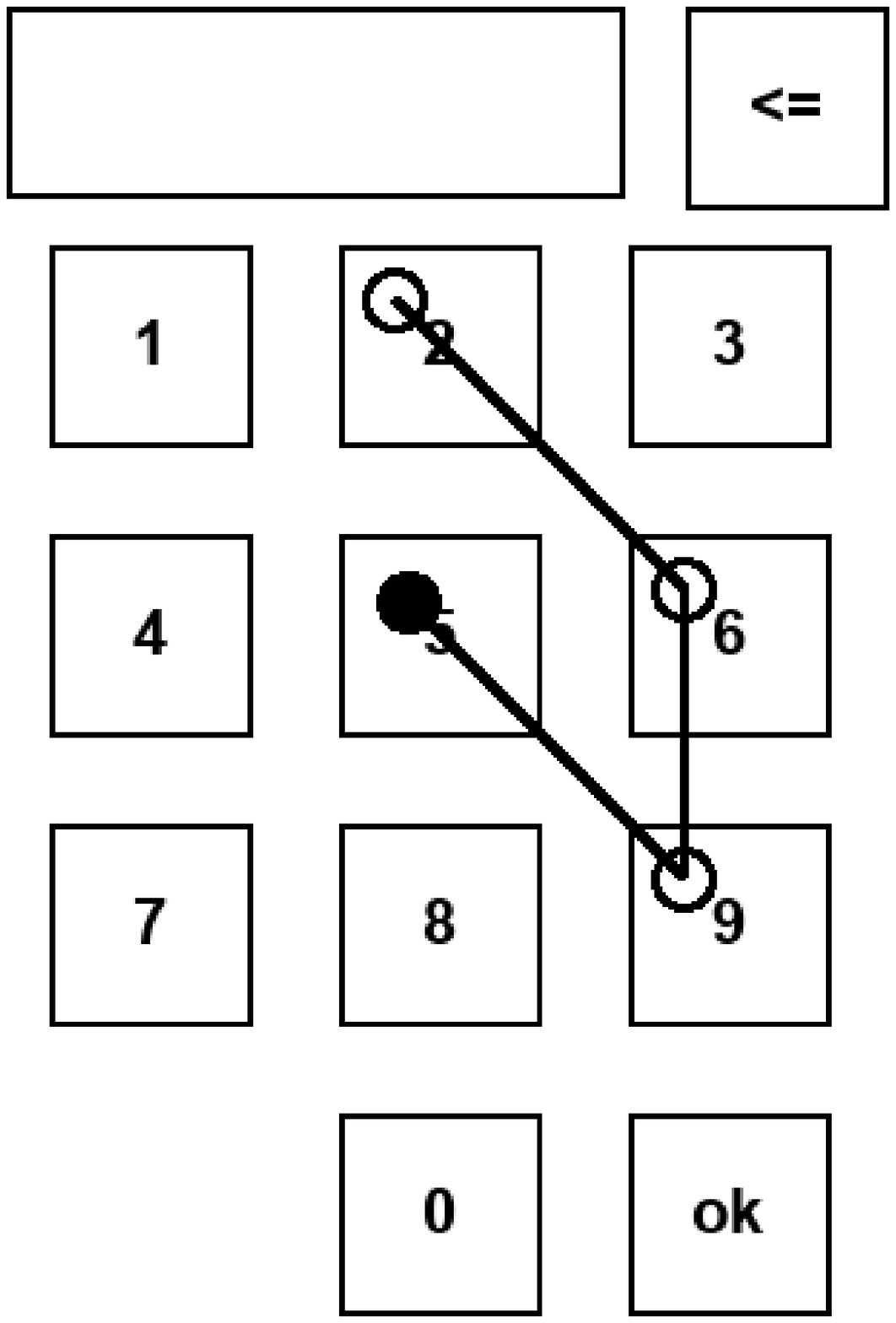}}&
    \fbox{\includegraphics[width=0.1\linewidth]{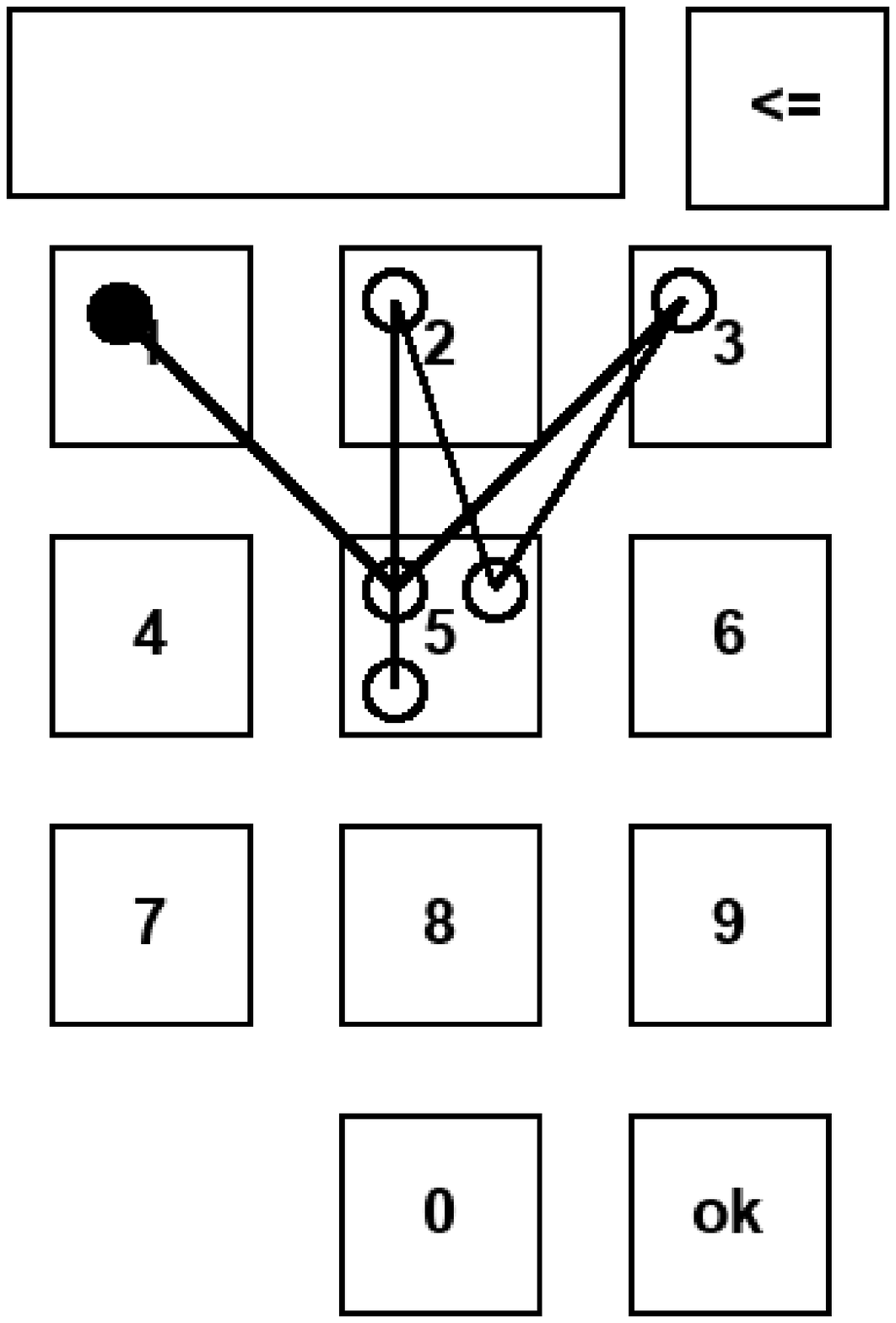}}\\
 5962 & 152525\\
 right & up/repat \\\\
  \end{tabular}

  \begin{tabular}{c c}
    \fbox{\includegraphics[width=0.1\linewidth]{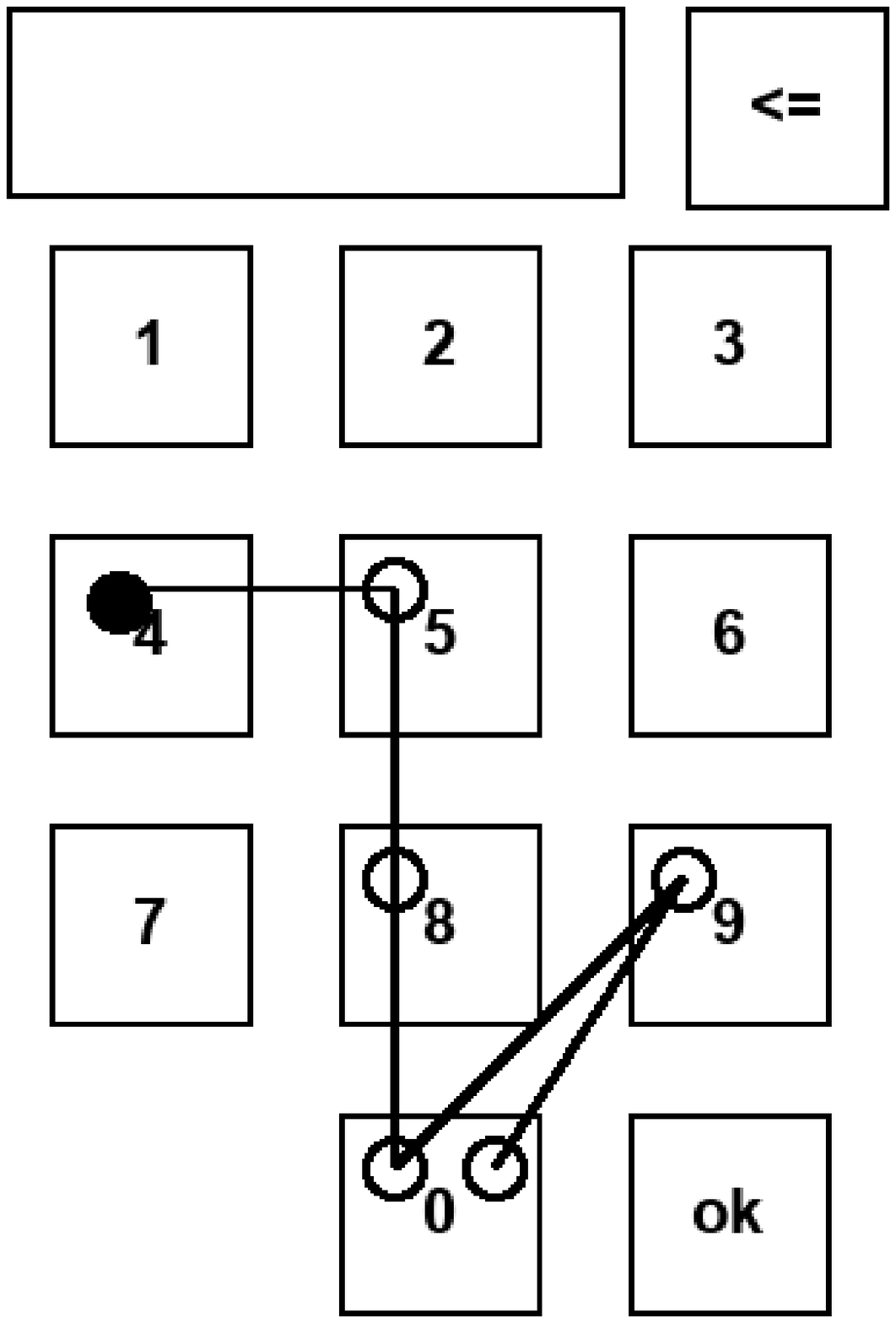}}&
    \fbox{\includegraphics[width=0.1\linewidth]{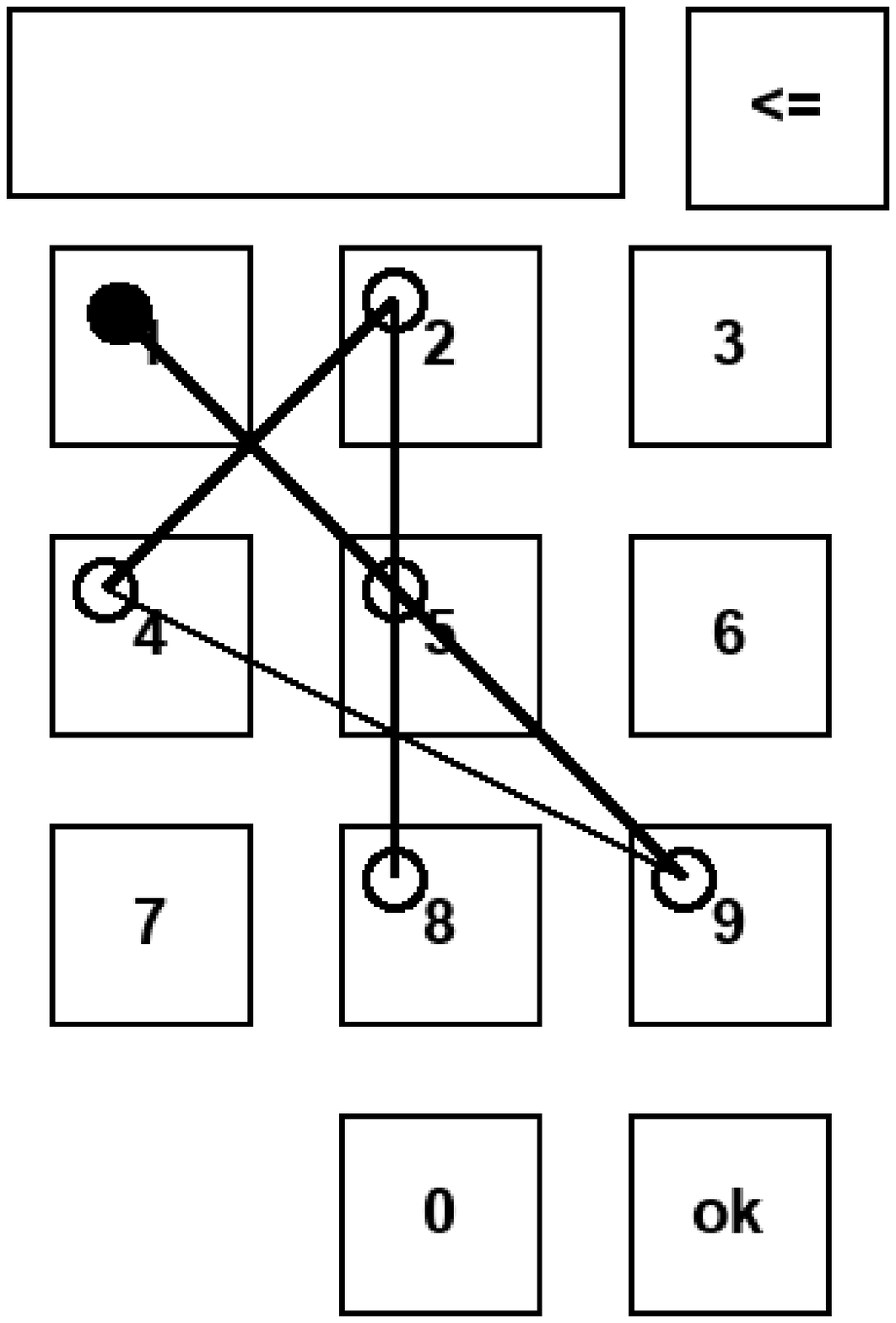}}\\
 458090 & 159428 \\
 down/repeat & neutral/cross/non-adj\\\\
  \end{tabular}

  \begin{tabular}{c c}
    \fbox{\includegraphics[width=0.1\linewidth]{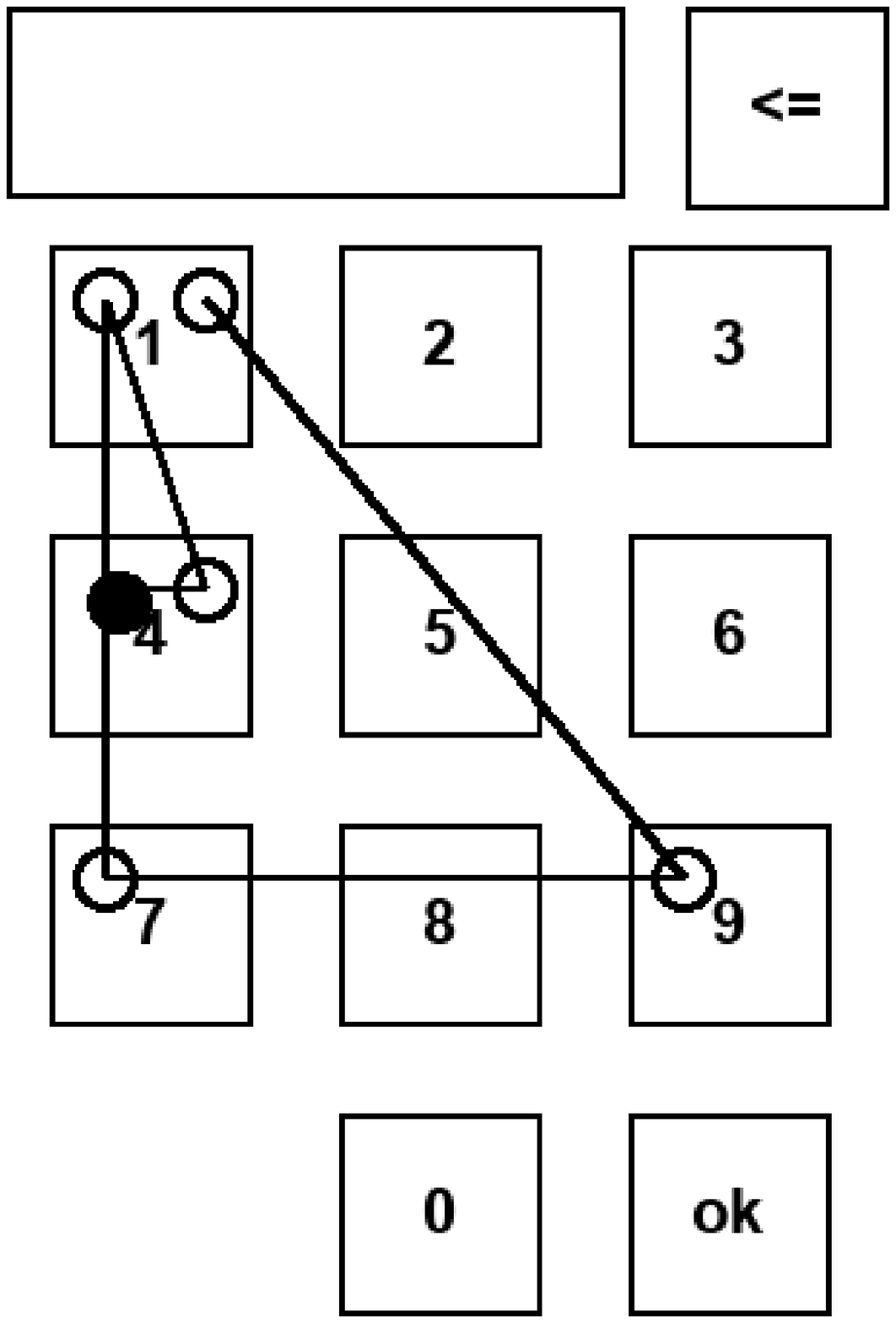}}&
    \fbox{\includegraphics[width=0.1\linewidth]{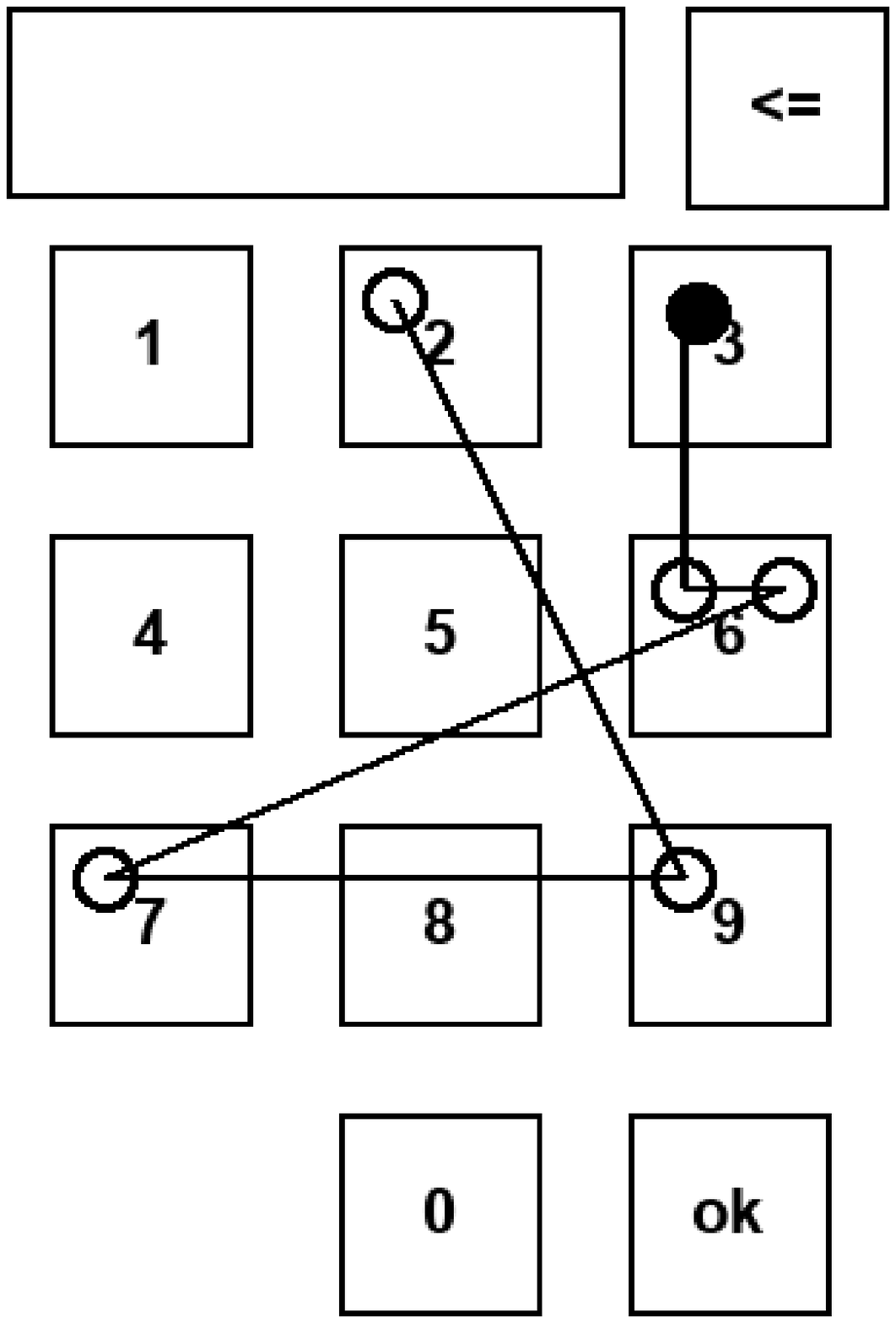}}\\
 441791 & 366792 \\
 left/kmove/repeat & right/repeat/kmove/cross\\
  \end{tabular}
\end{center}
